\documentclass[acmsmall,screen,nonacm]{acmart}

\usepackage{booktabs} %
\usepackage{subcaption}
\usepackage{tikz}
\usepackage{pgfplots}
\usepackage{pgfplotstable}
\usepackage{listings}
\usepackage{xcolor}
\usepackage{framed}
\usetikzlibrary{calc}
\usetikzlibrary{positioning}
\usetikzlibrary{arrows.meta}
\usetikzlibrary{fit}
\usetikzlibrary{backgrounds}
\usetikzlibrary{shadows}

\pgfplotsset{compat=1.15}
\usetikzlibrary{backgrounds,shapes.arrows,shapes.symbols,patterns,positioning,decorations.pathreplacing,calc,shapes.geometric,shapes.multipart,graphs,fit,arrows.meta}
\usepgfplotslibrary{groupplots}

\definecolor{tumblue}{RGB}{0,101,189}
\definecolor{dkgreen}{rgb}{0,0.6,0}
\definecolor{mauve}{rgb}{0.58,0,0.82}

\definecolor{tumsec1}{RGB}{0,82,147}
\definecolor{tumsec2}{RGB}{0,51,89}
\definecolor{tumbeige}{RGB}{218,215,203}
\definecolor{tumorange}{RGB}{227,114,34}
\definecolor{tumgreen}{RGB}{162,173,0}
\definecolor{tumlblue}{RGB}{152,198,234}
\definecolor{tummidblue}{RGB}{100,160,200}

\definecolor{TLlightblue}{RGB}{119,170,221}
\definecolor{TLlightcyan}{RGB}{153,221,255}
\definecolor{TLmint}{RGB}{68,187,153}
\definecolor{TLpear}{RGB}{187,204,51}
\definecolor{TLolive}{RGB}{170,170,0}
\definecolor{TLlightyellow}{RGB}{238,221,136}
\definecolor{TLorange}{RGB}{238,136,102}
\definecolor{TLpink}{RGB}{255,170,187}
\definecolor{TLpalegray}{RGB}{221,221,221}
\definecolor{Fuchsia}{RGB}{140,54,140}
\definecolor{TBblue}{RGB}{68,119,170}
\definecolor{TBcyan}{RGB}{102,204,238}
\definecolor{TBgreen}{RGB}{34,136,51}
\definecolor{TByellow}{RGB}{204,187,68}
\definecolor{TBred}{RGB}{238,102,119}
\definecolor{TBpurple}{RGB}{170,51,119}
\definecolor{TBgrey}{RGB}{187,187,187}

\AtBeginDocument{%
  }

\setcopyright{acmlicensed}
\copyrightyear{2018}
\acmYear{2018}
\acmDOI{XXXXXXX.XXXXXXX}

\acmConference[Conference acronym 'XX]{Make sure to enter the correct
  conference title from your rights confirmation emai}{June 03--05,
  2018}{Woodstock, NY}
\acmISBN{978-1-4503-XXXX-X/18/06}

\definecolor{dkgreen}{rgb}{0,0.6,0}
\definecolor{mauve}{rgb}{0.58,0,0.82}
\pgfplotsset{
  nodes near coords boundcheck/.style={
    visualization depends on={rawy \as \rawy},
    nodes near coords={\ifdim\rawy pt>\pgfkeysvalueof{/pgfplots/ymax}pt\pgfmathprintnumber\rawy\else\fi},
    nodes near coords style={font=\normalsize},
    restrict y to domain*={
    \pgfkeysvalueof{/pgfplots/ymin}:\pgfkeysvalueof{/pgfplots/ymax}
    },
  },
  nodes near coords boundcheckx/.style={
    visualization depends on={rawx \as \rawx},
    nodes near coords={\ifdim\rawx pt>\pgfkeysvalueof{/pgfplots/xmax}pt\pgfmathprintnumber\rawx\else\pgfmathprintnumber\rawx\fi},
    nodes near coords style={font=\scriptsize,/pgf/number format/.cd,precision=1},
    restrict x to domain*={
    \pgfkeysvalueof{/pgfplots/xmin}:\pgfkeysvalueof{/pgfplots/xmax}
    },
  },
}

\begin{document}

\title{TPDE: A Fast Adaptable Compiler Back-End Framework}

\author{Tobias Schwarz}
\email{tobias.schwarz@tum.de}
\affiliation{%
  \institution{Technical University of Munich}
  \country{Germany}
}
\author{Tobias Kamm}
\email{tobias.kamm@tum.de}
\affiliation{%
  \institution{Technical University of Munich}
  \country{Germany}
}
\author{Alexis Engelke}
\email{engelke@tum.de}
\affiliation{%
  \institution{Technical University of Munich}
  \country{Germany}
}

\begin{abstract}
Fast machine code generation is especially important for fast start-up just-in-time compilation, where the compilation time is part of the end-to-end latency. However, widely used compiler frameworks like LLVM do not prioritize fast compilation and require an extra IR translation step increasing latency even further; and rolling a custom code generator is a substantial engineering effort, especially when targeting multiple architectures.

Therefore, in this paper, we present TPDE, a compiler back-end framework that adapts to existing code representations in SSA form. Using an IR-specific adapter providing canonical access to IR data structures and a specification of the IR semantics, the framework performs one analysis pass and then performs the compilation in just a single pass, combining instruction selection, register allocation, and instruction encoding. The generated target instructions are primarily derived code written in high-level language through LLVM's Machine IR, easing portability to different architectures while enabling optimizations during code generation.

To show the generality of our framework, we build a new back-end for LLVM from scratch targeting x86-64 and AArch64. Performance results on SPECint 2017 show that we can compile LLVM-IR 8--24x faster than LLVM \texttt{-O0} while being on-par in terms of run-time performance. We also demonstrate the benefits of adapting to domain-specific IRs in JIT contexts, particularly WebAssembly and database query compilation, where avoiding the extra IR translation further reduces compilation latency.

\end{abstract}



\maketitle

\section{Introduction}

Just-in-time (JIT) compilation is a widely used technique for improving the performance of use cases ranging from efficient execution of dynamic languages like JavaScript~\cite{v8turbofan,webkit2016b3}, bytecode languages like WebAssembly~\cite{wasmtime,peach2020ewasm} or Java~\cite{paleczny_jvm01_java}, acceleration of database queries~\cite{neumann2011efficiently,VOILA,diaconu2013hekaton,FlounderIR}, to system emulation with binary translation~\cite{bellard2005qemu}.
A crucial part for a high-quality user experience is a low startup time, in which the input must be analyzed and transformed to machine code.
Within this trade-off of generating code fast and producing high-quality code, such runtime systems often use a multi-tiered compilation system, where a fast compiler for quick startup is typically paired with an optimizing compiler for achieving high execution performance. In addition to JIT compilation, low-latency compilation is also important for developer productivity to shorten compile--test cycles.

A very popular compiler framework is LLVM~\cite{lattner2004llvm}, which also has built-in support for JIT code execution. The framework not only features a high-quality optimizer and machine code generator, but also provides an unoptimized compilation pipeline that is substantially faster.
However, many JIT compilers have their own intermediate code representation (IR), requiring an extra translation to LLVM's IR, which increases latency.
In addition to that, the compilation times of LLVM are generally high, even in the unoptimized pipeline~\cite{engelke2024compile}. Thus, several systems have moved away from LLVM as their baseline compiler and built their custom back-end~\cite{kersten2021tidy,webkit2016b3,FlounderIR,v8turbofan,gruber2023bringing}.

However, rolling a custom back-end requires finding solutions for complex problems like register allocation and ABI implementation and is therefore a substantial effort to develop and maintain.
This effort increases further when porting the back-end to a new architecture, as many parts have to be adapted.
Moreover, such custom back-ends only support a single IR and are typically deeply embedded into their surrounding system, making it hard to reuse the code for other projects.

To address these problems, we present TPDE, a flexible compiler back-end framework focusing on fast compilation, where the generation of machine code including register allocation happens in just a single pass.
Instead of rolling a new custom IR, our framework can be flexibly adapted to existing IRs that represent code in Single Static Assignment (SSA) form.
To compile code with our framework, a user specifies two components: (a) an \emph{IR adapter}, through which the framework can query information about the IR like the successors of a basic block or the operands of an instruction; and (b) a set of \emph{instruction compilers}, functions which generate machine code for an instruction, essentially specifying the semantics.
The framework performs a liveness and loop analysis of the input IR and then steers the code generation, taking care of register allocation, spilling, and ABI handling for function calls and parameters. %
During code generation, the instruction compiler can call back into the framework for allocating temporary registers or for communicating register constraints.

To simplify writing instruction compilers and ease porting to different architectures, we provide a tool to generate snippets of target instruction sequences from LLVM's Machine IR, so that the semantics of to-be-generated code can be specified in a high-level language like C/C++ while maintaining the flexibility of assigning registers and performing low-level optimizations like use of more complex addressing modes. These snippet encoders can then be called by an instruction compiler, allowing an architecture-independent implementation of many instructions.

Using our framework, we implemented a compiler for LLVM-IR, fully independent of LLVM's existing code generation infrastructure, that is capable of compiling typical unoptimized code, for example, as generated by Clang for C and C++ programs.
This not only reduces the compile-time of ahead-of-time compilers like Clang or Rustc in total, but, more importantly, also systems that already use LLVM for JIT compilation (e.g., PostgreSQL~\cite{melnik2016postgres}, Clang-Repl~\cite{clang2025clangrepl}, Julia~\cite{bezanson2012julia}) can easily use our TPDE-based back-end as a significantly faster baseline compiler.
Our LLVM back-end targets x86-64 and AArch64 and consists of less than 8k lines of code.
Performance results evaluating the SPEC CPU2017 benchmarks show that our back-end is 8--24x faster than LLVM's compile-time-focused \texttt{-O0} pipeline while achieving similar run-time performance of the generated code ($\pm9\%$).

To show the flexibility of our framework and its benefits for JIT compilation, we also implement a back-end for Cranelift IR~\cite{byta2023cranelift,clifdoc} in the context of Wasmtime~\cite{wasmtime} to compile WebAssembly code, where our TPDE-based back-end was able to outperform Cranelift's fast compilation mode in compile-time and run-time performance.
We also implement a compiler for the Umbra database system~\cite{neumann2020umbra}, which uses a custom, domain-specific IR in SSA-form for compiling SQL queries to machine code. Here, our TPDE-based back-end is capable of being as fast as Umbra's specialized and highly-optimized direct emit back-end~\cite{kersten2021tidy,gruber2023bringing} while maintaining a similar performance on the generated code. %

The main contribution of this paper are:

\begin{itemize}
  \item A novel and highly efficient compiler framework which adapts to existing IRs in SSA form and only requires a specification of how to access IR data structures and the semantics of the IR instructions.
  \item An approach to extract code generation snippets from LLVM's Machine IR utilizing available instruction and data flow information for further optimizations and additionally significantly reducing the effort for porting a custom compiler to a different architecture.
  \item An implementation of a fast, single-pass code generation back-end for LLVM-IR targeting x86-64 and AArch64, which is 8--24x faster than the LLVM \texttt{-O0} pipeline while achieving similar code quality, making LLVM again a suitable candidate for baseline JIT compilation.
\end{itemize}

The remainder of this paper is structured as follows:
In \autoref{sec:challenges}, we revisit the general challenges of writing a single-pass machine code generator.
Next, \autoref{sec:tpde} describes the TPDE framework itself.
Afterwards, \autoref{sec:encgen} covers our approach to derive code generation snippets from a high-level language and their integration into TPDE.
We then show our implementation and benchmark results of our back-ends for LLVM-IR, WebAssembly, and Umbra IR in sections~\ref{sec:llvm}, \ref{sec:cranelift}, and~\ref{sec:umbra}, respectively.
Finally, \autoref{sec:rltdwork} covers related work and in \autoref{sec:summary} we summarize our findings.

\section{Challenges of Single-Pass Compilation}
\label{sec:challenges}
Typically, compilers perform instruction selection, register allocation, and machine code emission in separate stages. This allows each component to have a global view of the code and do non-local changes. For example, many instruction selection approaches consider the operands of an instruction for generating combined and therefore more efficient code sequences; and likewise, register allocation algorithms rely on information about the registers that are live in parallel and being able to insert spill code and register moves at arbitrary points.

When doing code generation in a single pass, where all three steps are done at once, code has to be generated in the order of the instructions. It is not easily possible to modify already generated code; doing so would be very costly and require extensive tracking of state, negating the performance benefits of combining these steps.
Thus, the instruction selector effectively cannot merge operands into one target instruction, as the code for the operands has already been generated; it can only consider merging IR instructions that come \emph{after} the current instruction.
Furthermore, the register allocator cannot know a priori how many registers are going to be used, because lowering an instruction might require additional temporary registers, possibly with constraints to specific registers.
These constraints rule out a large portion of existing instruction selection and register allocation algorithms; in general only local decisions are possible.

Even with single-pass compilation, some parts of the function prologue cannot be generated before compilation is complete; for example, the stack frame size is only known when the register allocator can finally determine how many stack spill slots are required. Therefore, the prologue needs to be generated in a way that it can be easily modified at the end.

\section{Framework}
\label{sec:tpde}
\begin{figure}
\resizebox{\linewidth}{!}{
\begin{tikzpicture}[block/.style={text width=3cm,draw,fill=white,align=center,minimum height=7ex}]
\node (ir) {IR};
\node[block,right=1cm of ir] (ana) {Analysis Pass \\ \small (Liveness, Loops)};
\node[block,right=1cm of ana] (cg) {Code Gen. Pass \\ \small (Regalloc, ABI, \dots)};
\node[block,right=1cm of cg] (jit) {In-Memory Mapping (JIT)};
\node[block,above=1ex of jit] (obj) {Object File Generation};
\node (fef) at ($(ana.north west)!.5!(cg.north east)+(0,0.5)$) {For each function};
\draw[-|] (fef) -- (fef -| ana.north west);
\draw[-|] (fef) -- (fef -| cg.north east);
\node[right=1cm of obj,align=left] (mc) {Object File};
\node[right=1cm of jit,align=left] (im) {In-Memory \\ Code};
\draw[dashed,-latex] (ir) -- (ana);
\draw[dashed,-latex] (ana) -- (cg);
\draw[dashed,-latex] (cg) -- (obj.west);
\draw[dashed,-latex] (cg) -- (jit);
\draw[dashed,-latex] (obj) -- (mc);
\draw[dashed,-latex] (jit) -- (im);

\begin{scope}[on background layer]
    \node[draw=black,fill=gray!10,inner sep=10pt,rounded corners=10pt,anchor=north west,fit=(ana)(cg)(fef)(obj),label={above}:TPDE Framework] (tpde) {};
\end{scope}

\node[block,minimum height=10ex,below=1.5cm of ana] (adap) {IR Adapter \\ \small (Access to IR data structures)};
\draw[-latex] (tpde.south -| adap) -- (adap);
\node[block,minimum height=10ex,below=1.5cm of cg] (ic) {Inst. Compilers \\ \small (ISel; semantics of IR Instructions)};
\draw[-latex] ($(cg.south -| ic)-(0.2,0)$) -- ($(ic.north)-(0.2,0)$);
\draw[latex-] ($(tpde.south -| ic)+(0.2,0)$) -- ($(ic.north)+(0.2,0)$);

\begin{scope}[on background layer]
    \node[draw=black,fill=gray!10,inner sep=10pt,rounded corners=10pt,anchor=north west,fit=(adap)(ic),label={below}:IR-Specific Parts] (iradap) {};
\end{scope}

\draw ($(ir.west |- iradap.south)+(0,-.75)$) coordinate (lw) -- ($(mc.east |- iradap.south)+(0,-.75)$) coordinate (le);
\node[above=1ex of le,anchor=south east] {Section 3};
\node[below=1ex of le,anchor=north east] {Section 4};

\node[align=center] (snipc) at ($(adap |- le)+(0,-1)$) {Snippet Code in \\ High-Level Language};
\node[block,right=1cm of snipc] (encgen) {Ahead-of-time \\ Snippet Prepare};
\node[right=1cm of encgen,
  double copy shadow={
    shadow xshift=0.5ex,
    shadow yshift=-0.5ex
  }, fill=white,draw,align=center] (snip) {Snippet \\ Encoders};
\draw[-latex,dashed]  (snipc) -- (encgen);
\draw[-latex,dashed]  (encgen) -- (snip);

\draw[-latex] (ic.east) -| ($(snip.north)+(-0.2,0)$);
\draw[-latex] ($(snip.north)+(0.2,0)$) -- ($(tpde.south -| snip.north)+(0.2,0)$);

\end{tikzpicture}
}
\caption{Overview of the TPDE compilation framework. The framework adapts to any IR in SSA form through an IR adapter, which exposes relevant IR properties in a canonical form, and instruction compilers, which provide the actual semantics for IR instructions. Instruction compilers can optionally make calls into instruction snippet encoders, which are generated ahead-of-time from a high-level language.}
\label{fig:overview}
\end{figure}
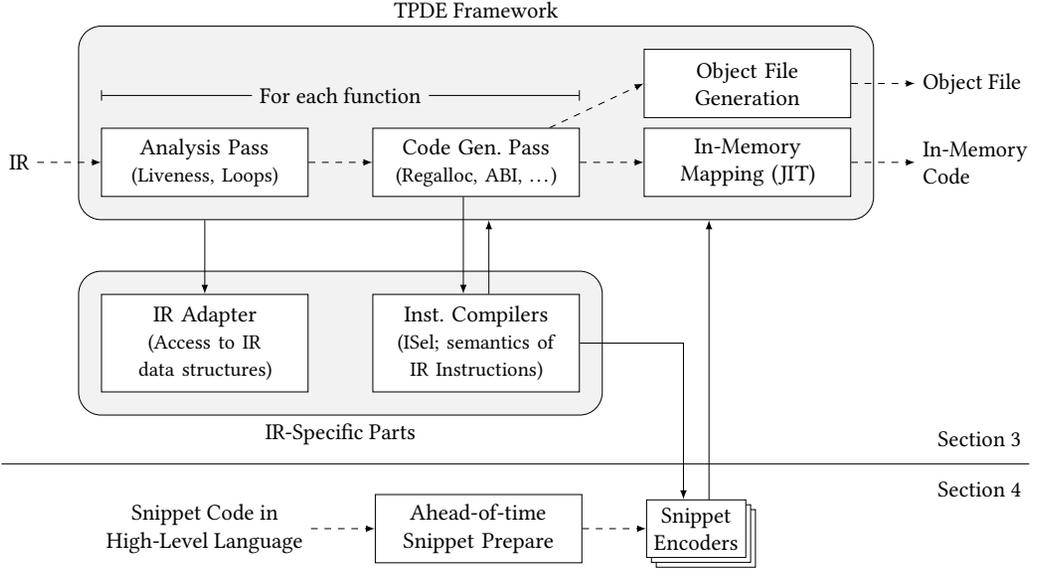

Our main goal is a very fast and reusable compiler framework.
In contrast to many existing systems~\cite{byta2023ircomp,lattner2004llvm,webkitb3ir}, however, we do not want to restrict ourselves to one specific IR which has to serve all use cases and requires a separate translation step from a previous code representation. Instead, we want to design the framework in a mostly IR-agnostic way that is capable of adapting to widely used IRs. By avoiding the IR translation step and additionally relying on code specialization through C++ templates to avoid an extra level of indirection between the framework and its user, we minimize the incurred performance cost of using a compiler framework instead of rolling a custom code generator.
We implemented this approach in our novel compiler framework TPDE\footnote{The code for the framework, snippet extraction and the LLVM back-end are available at \url{https://github.com/tpde2/tpde}.
};
\autoref{fig:overview} gives an overview over the architecture of the framework.

\subsection{Design}
\subsubsection{Separating IR-Dependent Components}
A primary objective while designing an IR-agnostic compiler framework was the separation of parts that tend to be highly dependent on the IR, e.g., instruction semantics and concrete data structures, and parts that tend to be less dependent on the IR and more reusable, e.g., register allocation and implementing ABI specifics.
Therefore, to adapt TPDE to an IR, a user needs to supply two components: The first component is an \emph{IR adapter}, through which the framework can query all needed information about the IR in a canonical way, like instruction operands or successors of a basic block. The second component is the \emph{instruction compiler}, which has to generate code for a single IR instruction. It can query the framework about the current location of its operands, require registers, possibly with constraints, emit instruction or data fragments, and communicate the location of the instruction results back to the framework.
For several other parts of the framework, like ABI handling, default implementations are provided, but can be overridden if required.

\subsubsection{IR Requirements}
\label{sec:irreq}
We designed our framework around IRs that are in strict SSA form, which is extremely popular among compilers, including the LLVM ecosystem~\cite{lattner2004llvm}, the MLIR ecosystem with its wide range of dialects~\cite{lattner2021mlir}, and many others~\cite{byta2023ircomp,webkitb3ir,neumann2020umbra,cytron1991efficiently}.
Requiring SSA form allows a more coarse-grained and therefore faster liveness analysis and the immutability of values simplifies tracking value states during code generation.
However, expensive construction of SSA form for variables is not required: mutable variables can be assigned to stack slots. Constructing this type of SSA from, e.g., expression trees is often cheap and requires no further analyses.

In our framework, we went with the notion of $\phi$-nodes, although in some IRs block arguments as an alternative formulation are becoming increasingly popular. We note that this is primarily a syntactic difference. Implementations using block arguments, however, tend to allow multi-edges between two blocks with different values in contrast to $\phi$-node-based implementations. TPDE currently does not support such multi-edges, but support could be added with reasonable effort.

Many IRs allow for instructions that produce multiple values, e.g., MLIR, values that consist of multiple components, e.g., LLVM values of struct or array type, or values that require multiple registers to be represented, e.g., 128-bit integers.
To capture all of these cases with one design, we model IR values generally as multi-part values. An IR adapter can then specify the number of parts for each IR value and, for each value part, its size and preferred register bank (typically general-purpose or floating-point/vector). In our framework, all value parts are treated separately for the purpose of register allocation.
Some IRs like LLVM allow constants to be used at any place where a value can be used. To cover this, a value part can also be a constant.

As the framework is not directly concerned with actual instruction semantics other than $\phi$-nodes, values are generally untyped and the only relevant property is their size, which is needed for moving values and spilling.
This keeps the framework flexible and prevents the need to constantly adjust the core for the wide range of types that are used in IRs.

\subsubsection{Two-Pass Approach}
As our key goal is fast compilation, we want to do code generation in just one single pass. Consequentially, instruction selection, register allocation, value spilling, and instruction encoding need to be done together in a linear pass over the IR.
We especially want to avoid creating another complete in-memory IR of the program, as this would not only have a significant performance impact, but also would impede the clean separation of IR and framework.

However, to achieve a reasonable code quality and size, liveness information about IR values is essential. As many values in SSA IRs tend to have only few users, often in the same basic block, liveness information allows to avoid generating unused spill code and to reuse registers as soon as their values are no longer needed.
Thus, we do a separate analysis pass to compute live ranges and the number of users of every value before the main code generation pass.

As a consequence of the single code generation pass, the compilation order of basic blocks has a substantial impact on length of the live ranges. Therefore, the analysis pass additionally determines the order in which blocks are compiled.

\subsubsection{Portability}
We also want our framework to support targeting different architectures (e.g., x86-64, AArch64) and platforms (e.g., Linux ELF, JIT).
Therefore, we separate the components of our compiler into parts that are independent of the architecture and platform, as well as parts that depend on the target architecture, the platform, or both.

For our goal of fast compilation and recomposability, a user of the framework can combine the architecture- and platform-specific parts as mixins for their final compiler configuration.
While this reduces the flexibility to dynamically recombine the architecture and platform, it avoids an indirection layer and thereby improves performance. Additionally, a user of the framework can easily supplement or replace many components of the framework to further tweak the functionality for their needs. We note that the set of relevant architecture--platform combinations that need to be available in the same binary is typically low.

\subsection{IR Adapter}
\noindent
The IR adapter is the only way for the framework to access the IR and therefore must expose all information that is required by the framework. The adapter is specified to the framework as a template parameter, enabling inlining of adapter methods and avoiding virtual function calls.
\autoref{fig:adapter} gives a complete list of the functionality that an IR adapter currently needs to provide.

\paragraph{Data Types and Initialization}
In the context of the IR adapter, the framework will refer to all functions, blocks, and values by reference data types defined by the IR adapter. However, for efficiency reasons, we recommend that only a single integer or pointer data type is used, as arrays of such reference types could grow unreasonably large otherwise.

\paragraph{Functions}
The adapter needs to provide a list of all functions that should end up in the symbol table, including both defined functions and declarations of external functions.
In addition to a symbol name, all functions also need to have a linkage (e.g., external, internal, weak).

Functions that have a definition need to be compiled. For this, the adapter also has to expose the function arguments and the basic blocks. Additionally, the adapter can also expose fixed-size stack variables, which will be allocated by the stack frame initialization of the framework.

\begin{figure}
\begin{minipage}{.33\linewidth}
\begin{lstlisting}[language=C]
General
  Function[] Functions
Function
  string SymbolName
  Linkage Linkage
  bool IsDefinition
  bool NeedUnwindInfo
  // For exception unwinding
  Function PersonalityFunc
  // E.g., static allocas
  StackVar[] StackVariables
  bool IsVariadic
  Argument[] Arguments
  Block[] BasicBlocks
\end{lstlisting}
\end{minipage}
\hfill
\begin{minipage}{.33\linewidth}
\begin{lstlisting}[language=C]
Block
  Block[] Successors
  Phi[] PHIs
  Instruction[] Instrs
  // 64-bit inline storage
  u64& AuxDataStorage
Value
  // Number to use as array
  // index for fast lookup
  u32 Number
  u32 PartCount
  u32 PartSize(partIdx)
  // Default register bank
  u8 PartRegBank(partIdx)
\end{lstlisting}
\end{minipage}
\hfill
\begin{minipage}{.3\linewidth}
\begin{lstlisting}[language=C]
Instruction : Value
  Value[] Operands
Phi : Value
  Block[] IncomingBlocks
  Value[] IncomingValues
Argument : Value
  u32 ByValSize
  u32 ByValAlign
  bool IsStructReturn
StackVar : Value
  u32 Size
  u32 Align
Constant : Value
  byte[] Data(partIdx)
\end{lstlisting}
\end{minipage}

\caption{Functionality required from an IR adapter. All instances are referred to by handles; \texttt{Value} has multiple sub-types. Basic blocks need to provide a 64-bit inline storage, values need a per-function unique number that is suitable as array index for access of data structures inside the framework.}
\label{fig:adapter}
\end{figure}

\paragraph{Basic Blocks}
A basic block consists of optional $\phi$-nodes and a list of instructions. As the analysis pass requires a control flow graph, it also needs to enumerate all possible successors.
We do not require an enumeration of predecessors, as not all IRs might have these readily available.
Furthermore, the adapter needs to expose 64 bits of storage per basic block. This way, the framework does not need to create hash tables for basic blocks itself, but can rely on more efficient ways to store data in basic blocks if provided by the IR. For IRs that already have auxiliary storage fields, the adapter can simply expose these, and for IRs like LLVM that number their blocks, the adapter can manage an array for this storage.

\paragraph{Values}
The framework distinguishes different types of values: instructions, $\phi$-nodes, arguments, stack slots, and constants. All values share that the adapter needs to expose the number of value parts (cf.~\autoref{sec:irreq}), their size, and their preferred register bank. The latter is used when the framework needs to copy values, for example when moving values to $\phi$-nodes.

Moreover, the adapter needs to expose a per-function-unique number of every non-constant value, which will be used as an array index. As the framework frequently needs to access per-value data structures, using an array gives a substantially better performance than a hash table.
Although many IRs do not provide such a numbering out-of-the-box, many IR value structures have auxiliary data fields, which can be used to store this number.
The sub-types of values have their usual properties. Arguments have some properties that are needed to correctly map them to target registers or stack slots, for example, to use the dedicated struct-return register on AArch64.

\paragraph{Initialization}
As the adapter implementation might want to compute a value numbering or allocate storage for the basic block auxiliary data, the adapter can implement two optional methods: prepare, which is called before the framework queries any information about a function, and finalize, which is called when associated data with the current function can be deallocated.

\subsection{Analysis Pass}
\label{sec:analysispass}
To determine when a value is no longer used and associated registers and stack slots can be reused, the code generator needs liveness information about arguments, $\phi$-nodes, and the results of all instructions.
This is particularly relevant for two-address architectures like x86-64, which often clobber one of their source operands --- without liveness information, the code generator would unconditionally add a copy, even at the last use of a value.

Various algorithms for liveness analysis have been proposed~\cite{kildall1973unified,boissinot2008fast,rastello2012sparse}, however, many of these provide more information than needed (resulting in avoidable computations) or require information that is not necessarily easily accessible, e.g., use lists.
To meet our goal of compile-time performance, instead of classical iterative algorithms, we use the algorithm proposed by Kohn et~al.~\cite{kohn2018adaptive}, which makes use of a loop forest instead and has a runtime linear in the number of instructions.
However, their approach performs a custom loop analysis which does not support irreducible loops.
As we want to support irreducible loops, we replace their loop analysis with the algorithm proposed by Wei et al.~\cite{wei2007loops}. Although this algorithm has a theoretical worst-case runtime of $\mathcal O(N\cdot E)$, the average runtime for typical CFGs is $\mathcal O(N+E)$ and it has lower constant factors than other loop finding algorithms that support irreducible CFGs, particularly as it avoids more expensive data structures like union-find.
Based on the loop analysis, the analysis pass also determines the order of basic blocks for later compilation.

Combining these, our analysis pass performs the following steps:

\begin{enumerate}
  \item Create a temporary numbering of all basic blocks, so that the loop analysis can store information about basic blocks in an array. The number of every basic block is stored in the auxiliary data field provided by the adapter.
  \item Identify loops following the algorithm by Wei et al.~\cite{wei2007loops}. For simplicity, we wrap the whole function in one single loop and build a loop tree similar to~\cite{kohn2018adaptive}.
  \item Compute block layout. We generally lay out blocks in reverse post-order with the minor addition that whenever a block is part of a loop, we place the whole loop together. This slightly shortens the live ranges of values whose liveness ends at the end of the loop.

  We store the final layout number of each block in the auxiliary data field of the adapter; from now on, basic blocks are referred to by this number. To mark visited blocks during the RPO traversal, we also use the auxiliary field instead of a separate set. We also track whether a block has more than one predecessor and store this in the auxiliary field.
  \item Compute liveness of arguments, instructions, and $\phi$-nodes using the second part of the algorithm from~\cite{kohn2018adaptive}. We also determine the number of users of every value. At the end, every value gets assigned a contiguous live range consisting of a start block number and an end block number as well as flag indicating whether the liveness ends within or at the end of the end block.
\end{enumerate}

This coarse-grained liveness information is sufficient for our purposes. We do not need a more precise beginning of the live range, as we generate code linearly in block-order and inside the block, the liveness implicitly begins at the definition of the value and, due to RPO block order, all uses come after the definition.
The information also allows to determine the end of the live range during code generation with sufficient accuracy. Later, the code generator will track the number of remaining uses for every value. When this value reaches zero and the live range ends \emph{within} the current block, the value becomes dead and the register can be reused immediately. Otherwise, even if there are no further uses, the value is still live due to a backedge, where the value must be live for the entire loop.

\subsection{Code Generation Pass}
The code generation pass compiles a function linearly into a code buffer. Blocks are compiled in the order determined in the previous analysis pass, instructions within the blocks are compiled in program order. Once code is written into the buffer, it might be fixed up later, e.g., for a jump to a later block, but is generally not moved or modified afterwards.
Therefore, the pass has to do instruction selection, register allocation, and machine code generation in a single step.

\subsubsection{Value Assignments}
For every live value, the framework stores an \emph{assignment}, which consists of a stack frame slot for spilling, the in-memory size, the number of remaining uses, and information about every value part. For each value part, we store the current register (if any), the size, and whether the stack slot already contains the correct value --- if not, the register is currently the only location of the value and must be spilled when evicted.

Additionally, value parts can be marked as \emph{trivially recomputable}, so that on eviction they do not get spilled. This is used, for example, for references to stack variables, which can be easily recomputed using frame pointer and offset.
Furthermore, a value can be \emph{locked}, implying that the value is currently not spillable. This is used to prevent eviction when reloading values into registers for use in an instruction.

For performance, we store all value assignments in an array, which is indexed by the value number provided by the IR adapter. As there are potentially many values needing assignments, we optimized the data structure for size: for single-part values, an assignment consists of just 16 bytes, with every additional part adding 2 bytes.

\subsubsection{Prologue and Epilogues}
\label{sec:tpde:pei}
When compiling a function, the first part that is generated is the prologue, which is done by the architecture-specific part of the framework.
As at this time neither the stack frame size nor the used callee-saved registers are known, sufficient space for saving all required registers is allocated. The final stack frame size and instructions to save callee-saved registers are patched into the prologue at the end; remaining instruction space is padded with no-ops.
Likewise, as it is not known whether a function uses dynamically-sized stack allocations, we always setup a frame pointer and do not use the stack pointer for referring to stack slots.

Similarly, when an epilogue is generated in the middle of the function, e.g., for functions with multiple return instructions, sufficient space for restoring callee-saved registers is allocated at first; only at the end, the actual instructions are filled in.

The framework also takes care of generating exception unwind information, which needs to provide information about the stack frame layout. As we always use a frame pointer, this information only needs to be updated after the prologue and before/after every epilogue. This information is written at the end when the stack frame layout is known. %

After the initial stack frame setup, the value assignments of the parameters are initialized with their respective locations in register or memory according to the current calling convention.

\subsubsection{Code Generation for Instructions}
To compile an instruction, the framework defers to the instruction compiler provided by the user of the framework, as only the user knows the actual semantics of their IR.
An instruction compiler typically does the following:
\begin{enumerate}
  \item Collect \emph{handles} to value parts that correspond to the instruction operands from the framework. The framework generally expects that all instruction operands are accessed; when a handle gets dropped, the remaining use count of the value will be decremented automatically. Having a handle also locks the value, preventing associated registers from getting spilled while the handle exists.
  \item Ensure that the required value parts are in registers. The framework will reload spilled values into registers, but due to value locking, existing handles and associated registers are guaranteed to remain valid.

  This is not strictly required, for architectures that can flexibly use memory operands, like x86-64, the framework can also supply a register--offset pair for spilled values.

  For value parts that need to be in special registers, e.g. due to instruction constraints, the framework allows to specify a set of feasible registers. However, such constrained values need to be loaded before unconstrained values to prevent locked values from blocking needed registers.
  \item Collect handles for the value parts that correspond to the instruction results and use these to allocate new registers. As some architectures like x86-64 often overwrite one of the operands, the framework provides methods to attempt reusing registers at the end of their liveness (cf. \autoref{sec:analysispass} for the exact conditions) or to generate a copy.
  \item Generate code for the actual semantics. As this might need extra registers, the framework can provide unevictable \emph{scratch registers} for this purpose.
  \item Notify framework about the location of the results. Should the result of a value part reside in such a scratch register at the end, the value assignment can also be updated to refer to that register, otherwise, scratch registers are released at the latest at the end of the instruction.
\end{enumerate}
\autoref{lst:example-add} shows an example for a simple instruction compiler.

\begin{lstlisting}[float,caption={Example instruction compiler for an addition instruction for x86-64. The framework will move values into registers and create a copy of the first operand if required.}, label={lst:example-add}, language=C++,captionpos=b]
void emit_add(IRValueRef inst) { // IRValueRef is defined by the Adapter
  ValuePartRef lhs_ref = val_ref(inst->getOperand(0), 0); // Handle for operand 0, part 0
  ValuePartRef rhs_ref = val_ref(inst->getOperand(1), 0); // Handle for operand 1, part 0
  // Get handle for result, communicate that lhs will be overwritten.
  // Will generate a copy into a new register if reuse is impossible.
  ValuePartRef res_ref = result_ref_will_overwrite(inst, 0, std::move(lhs_ref));
  AsmReg rhs_reg = val_as_reg(rhs_ref); // Force into register, might be spilled
  ASM(ADD64rr, res_ref.cur_reg(), rhs_reg); // Encode instruction, clobbers first operand
  set_value(res_ref, res_ref.cur_reg()); // Notify framework that register was modified
}
\end{lstlisting}

When generating a branch instruction, the instruction compiler needs to call into the framework to insert the necessary spill code. Afterwards, the operands for the branch instruction can be moved into registers. The branch, however, must be generated through the framework for handling $\phi$-nodes of the successor and for releasing registers whose liveness ends at the end of the block.

As steps (2)--(4) can be quite tedious to implement, especially when targeting multiple architectures, we provide an architecture-independent way to generate most parts of this from high-level languages; we describe this later in \autoref{sec:encgen}.

\subsubsection{Instruction Fusing}
A very important optimization is the fusion of adjacent instructions, for example, comparison with branch instructions and address calculation with load/store operations.
Thus, the framework supports fusing instructions from the same basic block: when an instruction is compiled and all users are in the same block, an instruction compiler decide to generate code for all the users and mark the original instruction as fused instead of providing result registers. This fusion is also supported transitively.

As we generate code in program order, only later instructions can be fused into their source instructions. Therefore, code for the later instructions will be generated at the point of the first instruction of the fusion and is only possible if the IR semantics permit such an instruction reordering. For simplicity and performance, instruction compilers will only want to look at immediately following instructions; the framework provides access to this list.

\subsubsection{Register Allocation}
Registers are allocated alongside code generation and therefore we only perform a strictly local, greedy approach. There is no possibility to change registers or insert spill code once the code is compiled.
When allocating a register and registers are available, the one with the lowest number is used. Otherwise, an arbitrary evictable register is chosen and spilled, in round-robin manner.
As a minor optimization, value parts that are used across multiple blocks inside the innermost loop can be assigned a \emph{fixed} register, which prevents the register from being spilled. This heuristic targets values defined in loops, especially $\phi$-nodes in the loop header, which typically contain loop induction variables. For these, frequent spilling and reloading would cause a more substantial performance degradation than reloading unchanged values from outside of the loop.
To avoid collisions with register constraints and calling conventions, only callee-saved registers that without a special purpose in the architecture are considered as fixed registers.

To further reduce compile time, we do not keep per-block register states, but only track the state at the current code generation point. This implies that for blocks where any predecessor is not immediately preceding in the block order, the register state is no longer available.
Therefore, when branching to a block with multiple predecessors or a block that does not immediately follow in layout order, we spill \emph{all} values that have no fixed register and are live at the entry of any of these successors. This way, all live values have a single, well-known location, which is either a fixed register or a stack slot.

Values for $\phi$-nodes are moved in their place after this spilling; critical edges are always split by inserting a separate block when there are values that need to be moved.

\begin{figure}
\begin{subfigure}{.4\linewidth}
\begin{lstlisting}[language=C,morekeywords={uint32\_t}]
uint32_t muli32(uint32_t a, uint32_t b) {
  return a * b;
}
\end{lstlisting}
\caption{Semantics expressed in C.}
\end{subfigure}
\hfill
\begin{subfigure}{.5\linewidth}
\begin{lstlisting}
bb.0 (%ir-block.2):
 liveins: $w0, $w1
 $w0 = MADDWrrr killed $w1, killed $w0, $wzr
 RET undef $lr, implicit killed $w0
\end{lstlisting}
\caption{Resulting LLVM Machine IR for AArch64.}
\end{subfigure}

\begin{subfigure}{\linewidth}
\begin{lstlisting}[language=C++]
void encode_muli32(AsmOperand param0, AsmOperand param1, ScratchReg &result0) {
  ScratchReg x0; // not yet allocated
  // $w0 = MADDWrrr killed $w1, killed $w0, $wzr
  AsmReg op1 = param1.as_reg_try_reuse(x0); // Ensure operands are in registers
  AsmReg op2 = param0.as_reg_try_reuse(x0); // w0/w1 marked as killed => try to reuse for result
  x0.alloc_from_bank(Config::GP_BANK); // If reuse was not possible, allocate a new register
  ASM(MADDw, x0.cur_reg, op1, op2, DA_ZR); // Actually encode the instruction
  param1.reset(); // killed, release any associated registers
  param0.reset(); // killed, release any associated registers
  // RET undef $lr, implicit killed $w0
  result0 = std::move(x0);
}
\end{lstlisting}
\caption{Generated snippet encoder. If any of the operands is in their last use, reuse the register for the result.}
\end{subfigure}

\caption{Overview of the snippet encoder generator targeting AArch64. Semantics are specified in a high-level language like C, which is compiled to the target-specific LLVM Machine IR. From there, we generate a function to generate the code, taking care of register allocation.}
\label{fig:encgenoverview}
\end{figure}

\section{Writing Instruction Compilers in High-Level Languages}
\label{sec:encgen}

While the framework as described in \autoref{sec:tpde} provides an abstraction for managing registers, it requires an explicit specification of the target instructions to be used for the operations. Although this allows for a very high flexibility for the code that is generated, writing such instruction compilers is often tedious and error-prone, especially for more complex instruction sequences. Moreover, when porting to a different architecture, the entire process has to be repeated.

To address these problems, we provide a way for writing snippets in a high-level language like C/C++. We compile these functions to LLVM-IR and from there further to LLVM's target-specific Machine IR (MIR), which contains not just information about the used instructions, but also about data flow dependencies, register usage, and register constraints. From the Machine IR of a function, we generate a \emph{snippet encoder} which, when called, dynamically adjusts the instruction sequence for the actual operands and available registers and then emits the resulting machine code, moving values into registers and allocating scratch registers as needed.

We implemented this approach to generate such snippet encoders for x86-64 and AArch64 as a supplementary tool to the main TPDE framework.
\autoref{fig:encgenoverview} gives an overview on the input and output of the tool.

\subsection{Generating Snippet Encoders}
As we rely on LLVM's Machine IR to extract the target instruction sequences, the input needs to be written in a language that can be compiled to LLVM-IR, for example, C or C++ using Clang.
After applying optimizations, we run the regular back-end pipeline until just before the actual machine code is emitted. At this point, the MIR only consists of encodeable instructions using solely physical registers. (See \autoref{sec:encgen:mirstoppass} for a discussion on alternatives.)

\subsubsection{Function Signature}
At this point, we first derive the function signature of the snippet encoder.
The function generally takes one parameter per input register followed by one parameter per output register.
For input parameters, we introduce a new type \texttt{AsmOperand}, which is a union type storing a value part handle or a scratch register. This increases the flexibility of the function; we will expand this later to enable further optimizations.
Output parameters are scratch registers, which are allocated inside the snippet encoder when required.

Input and output registers are mapped to parameters and return values in the original function according to the used calling convention. Hence, this is typically a one-to-one mapping, except for large values like 128-bit integers that split over two registers.
We currently only support functions where all parameters and return values are passed in registers. If the number of registers provided by the default calling convention is not sufficient, an alternative calling convention with more parameter or result registers can be used, e.g., regcall on x86-64.

\subsubsection{Tracking Registers}
\label{sec:encgen:regs}
At the beginning of the encoder, we allocate all constrained registers that are used throughout the function, e.g., for instructions that require a value in a specific register like the x86-64 division. We do this early to prevent later allocations from blocking such registers --- otherwise, a scratch register allocated later might happen to block this specific register, necessitating an inspection of all scratch registers at that point.

It can happen that no such register is currently available, for example, when a specific register is required but is currently fixed, e.g. due to use in a value handle. In this case, we forcefully move the value to a new register, updating references to the register in the input parameters. At the end of the function, we move the value back into its original location.
This simple strategy is possible, because the snippet encoder will not refer to IR values that are not supplied as parameters.

For all other used registers, we prepare one scratch register each, but allocate these lazily, as no constraints need to be satisfied. When generating code, we map every physical register in the MIR to either the corresponding scratch register or, in case the register is an input parameter and not yet overwritten, the \texttt{AsmOperand}.

We currently only support functions that do not use or modify the stack or frame pointer, as these are used by the generated function. However, handling such functions is generally possible by translating stack frame allocations of the function to use TPDE's infrastructure for stack allocations. We also note that such an extra stack frame setup is likely unwanted, as it would degrade compile-time and run-time performance. Instead of inlining such complex functions, generating a call to a runtime library seems generally preferable.

\subsubsection{Handling and Emitting Instructions}
For the main part of the snippet encoder, we handle the MIR instructions in order and generate appropriate code. %

First, we need to ensure that all registers used by the instruction are in encodeable registers. \texttt{AsmOperand}s that are not yet in a register need to be materialized, e.g. materializing the constant or by loading the value from the stack.
Some instructions clobber some of their input operands (\emph{tied} operands, e.g., x86-64 \texttt{add rax, rcx} reads and writes \texttt{rax}). If in such a case an input operand cannot be clobbered, because the value is still needed, we copy the value into a new output register.
As we do not change the order of instructions, we ignore certain implicit registers like flags or floating-point control registers.

Second, we need to make sure that all output registers are allocated. When the instruction has an \texttt{AsmOperand} in their last use as input, we try to reuse the register. Otherwise, we ensure that the scratch register associated with the physical register in the MIR is allocated.

Once all input and output operands are known, we generate a call to the assembler for encoding the resulting instruction.
As we do not use the LLVM-MC assembler for encoding instructions due to its subpar performance, we have to map the LLVM mnemonics to the ones of our assembler. While LLVM's mnemonics do follow a naming scheme, this is not very consistent and there are some exceptions. This required some manual effort to create such a mapping.
When an instruction refers to a constant from the constant pool of the MIR function, we add the constant to the read-only data section and emit a relocation.
Afterwards, for output registers of the instruction, we update our register mapping to point to their associated scratch register.

\subsubsection{Control Flow}
To increase the scope of supported functions, we also support MIR functions with multiple basic blocks. This can happen even for seemingly straight-forward operations, for example, when converting a 64-bit unsigned integer to a floating-point value on x86-64.

At the end of the first basic block, we materialize all \texttt{AsmOperand}s into scratch registers. From this point onward, we simply reproduce the instruction sequence generated by LLVM without any further optimizations, except that the used registers can differ.
Return instructions are generally replaced with a jump to the end of the generated code. This jump is omitted if there are no instructions to be jumped over, e.g., when the function consists of a single basic block with ending with a return.
Indirect jumps are currently not implemented; however, there should be no structural problems in adding support for these. Function calls require a stack frame, which is currently not supported (see \autoref{sec:encgen:regs}).

\subsection{Optimizing for Non-Register Operands}
\label{sec:encgen:asmop}
When embedding the code as part of the compilation process, values might be spilled to the stack or can be constants. Depending on the target architecture, such values can sometimes be encoded as memory or immediate operand. Furthermore, expressions like register with offset can often be encoded directly into memory operands, avoiding an extra instruction.

To enable the use of immediate encodings and more complex addressing modes, input operands (\texttt{AsmOperand}) cannot only be single registers, but also be a value handle, referring to either a register or a stack slot, a modifiable scratch register, a non-modifiable raw register, a constant, or a simple expression of the form $base+scale*index+offset$, where $base$ and $index$ are either scratch or raw registers. Expressions can also describe stack variables using the frame pointer as base.

When encoding an instruction with an \texttt{AsmOperand} input, the encoder checks whether the operand can be merged into the instruction. Otherwise, the operand gets materialized into a register. Depending on the available encodings of the instruction, the following variants are checked:
\begin{itemize}
  \item Replacing a register with an immediate operand for constants.
  \item Merging expressions into address operands. This also takes into account that the LLVM-generated instruction may additionally use its own offset; the combination of all address components must be encodeable.
  \item x86-64: using a memory operand for spilled IR values.
\end{itemize}

Merging expressions into memory operands has a large impact on the code size and performance for programs that frequently access stack variables --- otherwise, the rather simple address computation (frame pointer with constant offset) would result in a separate instruction.

\subsection{Omitting Register Moves}
The instruction sequence generated by LLVM is optimized as a complete function with fixed registers for inputs/outputs as specified by the calling convention and therefore often contains instructions to move values out of parameter or into result registers.
In our use of the code, however, we do not need to adhere to a calling convention, making such moves often avoidable.

Therefore, instead of generating move instructions that are known to have no other side effects used by the program (e.g., no implicit zero-extension), we mark the destination register as alias for the source register and do not allocate a separate register.
When the alias is used as operand, the actual source of the value is used instead. This allows doing the previously described optimizations even after a move.
However, care must be taken when the aliased register or the source register is overwritten.
In that case, an actual copy needs to be materialized.

\subsection{Discussion}

\subsubsection{Portability}

As MIR is a mostly target-independent code representation, porting the approach to a new architecture is possible with comparably low effort.
After describing the mapping of LLVM machine registers to TPDE register numbers, the only requirement for basic functionality is the description of how specific instructions can be encoded, which boils down to emitting corresponding assembler calls.

Encoding optimizations, however, are naturally very target-specific and therefore need to be implemented depending on encoding options provided by the target architecture.
The framework itself is very flexible: for every instruction, it queries the target-specific code for a list of all possible encoding candidates and their conditions. Conditions can be attached to every input register in form of arbitrary C++ code, which typically consist of calling a helper function, for example, ``is $reg+32$ encodeable as 32-bit immediate''.
The first candidate where all conditions are met used.

\begin{figure}
\begin{subfigure}{.42\linewidth}
\begin{lstlisting}[language=C,morekeywords={\_\_int128}]
typedef unsigned __int128 u128;
u128 shli128(u128 val, int amt) {
  return val << amt;
}
\end{lstlisting}
\begin{lstlisting}[language=C]
shli128:
  lsr     x8, x0, #1
  mvn     w9, <amt>
  lsl     x10, x1, <amt>
  mov     x2, <amt> // must materialize, tst
  tst     x2, #0x40 // cannot have two imms
  lsr     x8, x8, x9
  lsl     x9, x0, <amt>
  orr     x8, x10, x8
  csel    x0, xzr, x9, ne // amt >= 64?
  csel    x1, x9, x8, ne // amt >= 64?
  ret
\end{lstlisting}
\caption{Generic snippet and machine code generated by LLVM. Replacing the shift amount with a constant results in unneeded instructions.}
\label{fig:shiftsnippets:a}
\end{subfigure}
\hfill
\begin{subfigure}{.55\linewidth}
\begin{lstlisting}[language=C,morekeywords={\_\_int128,uint64\_t}]
typedef unsigned __int128 u128;
u128 shli128_lt64(u128 a, int amt, int iamt) {
  uint64_t lo = (uint64_t)a << amt;
  u128 hi0 = (uint64_t)a >> iamt; // iamt = 64-amt
  u128 hi1 = (uint64_t)(a >> 64) << amt;
  return (hi0 | hi1) << 64 | lo;
}
u128 shli128_ge64(u128 a, unsigned amt) {
  return a << 64+(amt%
}
\end{lstlisting}
\begin{minipage}{.45\linewidth}
\begin{lstlisting}
shli128_lt64:
  lsl     x8, x1, <amt>
  lsr     x9, x0, <iamt>
  lsl     x0, x0, <amt>
  orr     x1, x9, x8
  ret
\end{lstlisting}
\end{minipage}
\hfill
\begin{minipage}{.45\linewidth}
\begin{lstlisting}
shli128_ge64:
  lsl     x1, x0, <amt>
  mov     x0, xzr
  ret
\end{lstlisting}
\end{minipage}
\caption{Snippets optimized for constant shifts. Inserting the constant shift amount results in more efficient code without unneeded instructions; \texttt{iamt} is computed by the caller on the constant shift amount.}
\label{fig:shiftsnippets:b}
\end{subfigure}

\caption{Although constant operands are folded into snippets, instructions with all-constant inputs are not eliminated. Providing separate snippets for constants can significantly improve the generated code for some operations. (Machine IR lowered to assembly for readability.)}
\label{fig:shiftsnippets}
\end{figure}

\subsubsection{Writing Optimizable Snippets}
Our optimization for constant operands just attempts to encode the operand as immediate into the instruction. However, no constant-folding of instructions with all-constant inputs is performed --- this would require implementing the semantics for many architecture-specific instructions. This can result in several unneeded instructions, as exemplified in \autoref{fig:shiftsnippets:a}.
In some cases, providing additional snippets for specific value ranges can substantially improve the quality of the generated code, as shown in \autoref{fig:shiftsnippets:b}. Depending on whether the operand is a constant, the instruction compiler can either call the generic snippet or the variants optimized for specific ranges.

\subsubsection{Which Machine IR Stage?}
\label{sec:encgen:mirstoppass}
In the described approach, we use the Machine IR of a function in its latest stage after register allocation.
At this point, all instructions use encodeable physical registers and all LLVM-internal pseudo-instructions are lowered.
Although we have to eliminate unneeded moves inserted by the register allocator, we do not have to spill any registers but can be sure that sufficient registers are available. %

An alternative approach would be to use the MIR before register allocation, either in SSA-form or after $\phi$-node elimination. We previously implemented an approach that used MIR in SSA-form, but moved on to our current approach, which greatly simplified the implementation:
First, tracking the mix of virtual and physical registers substantially increases complexity when generating the code for allocating registers.
Second, at this stage, LLVM employs pseudo-instructions for several operations, e.g., zeroing a register. We would need to expand these manually, as the normally responsible LLVM pass requires allocated registers.
Third, when using MIR in SSA-form, we need to lower $\phi$-nodes ourselves when handling programs with control flow. Although the framework provides an implementation, it is designed for handling IR values and would require substantial changes for other types of values.

We also considered extracting the machine code sequence from generated object files. While this would make the process independent of the rather unstable LLVM API and allow using other compilers, it would require substantial effort to reconstruct information that is readily available in MIR, including data flow dependencies, register constraints, stack frame layout, and references to constant pool entries.

\section{Case Study: Compiling LLVM-IR}
\label{sec:llvm}

To show the generality of our framework, we build a fast baseline compiler for LLVM-IR targeting x86-64 and AArch64, most notably without using any of LLVM's existing code generation infrastructure.
As our goal is a baseline compiler, we limit ourselves to IR constructs that are typically found in unoptimized code, e.g., as produced by Clang on typical C++ programs.
We therefore exclude uncommon data types like integers larger than 128 bits, floating-point types other than \texttt{float}/\texttt{double}, and vector types.
We also currently do not support inline assembly and other rarely used features like garbage collection support or computed \texttt{goto}; however, we note that implementing these with our framework is structurally possible.
\subsection{Implementation}
\label{sec:llvm:structure}
\subsubsection{IR Adapter}
Many of the functions of the IR adapter translate naturally to function calls on LLVM's data structures. Nonetheless, we need to do a preparation pass over the IR of the function before the analysis pass to build some lookup data structures and to legalize some operations that would be difficult to handle later.

We first number each global value, block,
and instruction with linearly growing indices, which are then exposed to the framework as value/block references.
We store the mapping from number to value in an array, together with precomputed and cached type and part count.

We also convert constant expressions into normal instructions to simplify handling of constants in the code generation pass, especially when they occur in $\phi$-nodes. Additionally, LLVM supports access to thread-local variables at arbitrary places, which is difficult to implement as, depending on the ABI, constructing the actual address can involve an external function call. Hence, we rewrite all accesses to thread-local variables to explicitly use the \texttt{llvm.threadlocal.address} intrinsic.

\subsubsection{Compilation}
First, before compiling any functions, global variables and aliases are transformed into corresponding symbols and chunks in the data sections, generating relocations as appropriate. As the framework provides abstractions for the most common relocations (e.g., absolute 64-bit address), this code is also largely portable.

The implementations of many LLVM-IR instructions are architecture-independent by heavily relying on snippet encoders as described in \autoref{sec:encgen}.
The only exceptions are (a) calls/returns, for which the calling convention needs to be considered, (b) branches, (c) integer comparisons, and (d) target- or ABI-specific intrinsics like access to varargs.
Integer comparisons very often appear in combination with conditional branches. Fusing these pairs is very important for performance to generate typical compare--branch machine instructions instead of a conditional set followed by a branch if non-zero. As the snippet encoders currently cannot handle code for branching to different basic blocks, branches and integer comparisons are written in architecture-specific code.

Apart from these cases, the instruction compilers are a mostly straight-forward implementation of LLVM-IR semantics. Supporting arbitrary-bit-width integers (up to 128 bit) and aggregate values like structs, however, increases implementation complexity of related instructions due to zero/sign extension or handling multi-part values. Some operations and intrinsics need to generate calls into the standard library; as the snippet encoders currently cannot generate function calls, these must be generated explicitly for now.

\subsubsection{Code Complexity}
Although lines of code is by no means an accurate metric to measure code complexity, it still gives a rough estimate.
In total, our LLVM compiler consists of 7.5k lines of code excluding blank and comments, of which about 1.7k
are architecture-specific for x86-64 and AArch64 combined.
The approach of abstracting target-specific instructions with encoding snippets not only makes the implementation significantly easier to read, write and maintain, but also greatly reduces the effort of porting to a different architecture.
Adding support for targeting AArch64 was merely a matter of days and most of the code consisted of comparably simple logic only.
In a previous implementation without encoding snippets, we needed roughly 12.8k lines of code, of which around 11.6k were target specific. This shows that, although not required by the core framework, snippet encoders substantially reduce the amount of required code and increase portability between architectures.

\subsection{Performance Evaluation}
\label{sec:llvm:eval}
\subsubsection{Setup}
We evaluate the performance of our implementation by measuring the back-end compile-time and the run-time performance of LLVM-IR generated by Clang.
We use code produced with \texttt{-O0}, where all variables are stack-allocated, the IR has only very few $\phi$-nodes, and SSA values have short live ranges, as well as code produced with \texttt{-O1}, where variables are in SSA form.
We embedded our LLVM back-end into Clang and compare it against the LLVM \texttt{-O0} back-end, which focuses on fast compilation\footnote{The LLVM \texttt{-O0} back-end runs substantially fewer passes and uses a faster instruction selector and register allocator.}, and the LLVM \texttt{-O1} back-end. Additionally, we compare against the copy-and-patch-based LLVM-IR compiler from \cite{drescher2024fast} in an updated version that also supports C++ exceptions and can compile LLVM-IR directly without the overhead of MLIR. However, this compiler only supports x86-64 and only works with the unoptimized input IR; the compiler crashes on benchmark 623.xalanc.

As benchmark programs, we use the SPEC CPU2017 integer benchmarks on x86-64 and AArch64.
Our x86-64 machine is an Intel Xeon Gold 6430
with 256\,GiB of RAM running Linux 6.8.0;
Our AArch64 machine is an Apple M1 with 16\,GiB of memory, using only the four performance cores,
running Asahi Linux 6.11.9.
We use Clang/LLVM version 19.1.3.

\begin{figure}
\centering
\begin{subfigure}[b]{0.49\textwidth}
\resizebox{\textwidth}{!}{
  \begin{tikzpicture}
    \begin{axis}[
      ybar=0pt,
      ymin=0,
      ymax=26,
      ymajorgrids=true,
      ytick distance=4,
      bar width=7pt,
      symbolic x coords = {
        600.perl,
        602.gcc,
        605.mcf,
        620.omnetpp,
        623.xalanc,
        625.x264,
        631.deepsjeng,
        641.leela,
        657.xz,
        geomean
      },
      enlarge x limits=0.05,
      xtick=data,
      xtick pos=bottom,
      cycle list={%
        TLlightyellow!30!black,fill=TLlightyellow,mark=none,postaction={pattern=north east lines}\\%
        TLorange!30!black,fill=TLorange,mark=none\\%
        TLlightblue!30!black,fill=TLlightblue,mark=none\\%
        TLpink!30!black,fill=TLpink,mark=none\\%
        TLlightyellow!30!black,fill=TLlightyellow,mark=none\\%
        TLlightcyan!30!black,fill=TLlightcyan,mark=none\\%
        TLolive!30!black,fill=TLolive,mark=none\\%
        TLmint!30!black,fill=TLmint,mark=none\\%
        TLpear!30!black,fill=TLpear,mark=none\\%
        TLpalegray!30!black,fill=TLpalegray,mark=none\\%
        TLlightblue!30!black,fill=TLlightblue,mark=none\\%
      },
      xticklabel style={rotate=30,anchor=east,align=center,font=\LARGE},
      yticklabel style={font=\Large},
      legend style={font=\large,legend columns=-1,
        at={(1,1)},
        anchor=north east,},
      legend image code/.code={
        \draw [#1] (0cm,-0.1cm) rectangle (0.1cm,0.15cm); },
      width=12cm,
      height=5.5cm,
      nodes near coords boundcheck,
      ]

      \addplot coordinates {
        (600.perl,15.188524510612693)
        (602.gcc,16.888181750906533)
        (605.mcf,29.955893163440333)
        (620.omnetpp,30.257778521244564)
        (623.xalanc,0)
        (625.x264,15.227611203818341)
        (631.deepsjeng,6.099699616272811)
        (641.leela,27.264034817778175)
        (657.xz,23.97637976195947)
        (geomean,18.565919785254856)
      };
      \addplot table[x=benchmark,y expr=\thisrow{x86_64-clang}/\thisrow{x86_64-tpde},col sep=comma] {bench-res/parsed-spec-ct};
      \addplot table[x=benchmark,y expr=\thisrow{aarch64-clang}/\thisrow{aarch64-tpde},col sep=comma] {bench-res/parsed-spec-ct};

      \coordinate (A) at (axis cs:geomean,1);
      \coordinate (O1) at (rel axis cs:0,0);
      \coordinate (O2) at (rel axis cs:1,0);

    \end{axis}
  \end{tikzpicture}
}
  \caption{Compile-time speedup}
  \label{fig:llvm-ct}
\end{subfigure}
\begin{subfigure}[b]{0.49\textwidth}
\resizebox{\textwidth}{!}{
  \begin{tikzpicture}
    \begin{axis}[
      ybar=0pt,
      ymax=1.4,
      ymin=0,
      ymajorgrids=true,
      ytick distance=0.2,
      bar width=7pt,
      symbolic x coords = {
        600.perl,
        602.gcc,
        605.mcf,
        620.omnetpp,
        623.xalanc,
        625.x264,
        631.deepsjeng,
        641.leela,
        657.xz,
        geomean
      },
      enlarge x limits=0.05,
      xtick=data,
      xtick pos=bottom,
      cycle list={%
        TLlightyellow!30!black,fill=TLlightyellow,mark=none,postaction={pattern=north east lines}\\%
        TLorange!30!black,fill=TLorange,mark=none\\%
        TLlightblue!30!black,fill=TLlightblue,mark=none\\%
        TLpink!30!black,fill=TLpink,mark=none\\%
        TLlightyellow!30!black,fill=TLlightyellow,mark=none\\%
        TLlightcyan!30!black,fill=TLlightcyan,mark=none\\%
        TLolive!30!black,fill=TLolive,mark=none\\%
        TLmint!30!black,fill=TLmint,mark=none\\%
        TLpear!30!black,fill=TLpear,mark=none\\%
        TLpalegray!30!black,fill=TLpalegray,mark=none\\%
        TLlightblue!30!black,fill=TLlightblue,mark=none\\%
      },
      xticklabel style={rotate=30,anchor=east,align=center,font=\LARGE},
      yticklabel style={font=\Large},
      legend style={font=\Large,legend columns=-1,
        at={(1,1)},
        anchor=north east,},
      legend image code/.code={
        \draw [#1] (0cm,-0.1cm) rectangle (0.2cm,0.25cm); },
      width=12cm,
      height=5.5cm,
      ]

      \addplot table[x=benchmark,y expr={\thisrow{x86_64-copypatch}>0?\thisrow{x86_64-clang}/\thisrow{x86_64-copypatch}:0},col sep=comma] {bench-res/parsed-spec-rt};
      \addplot table[x=benchmark,y expr=\thisrow{x86_64-clang}/\thisrow{x86_64-tpde},col sep=comma] {bench-res/parsed-spec-rt};
      \addplot table[x=benchmark,y expr=\thisrow{aarch64-clang}/\thisrow{aarch64-tpde},col sep=comma] {bench-res/parsed-spec-rt};

      \coordinate (A) at (axis cs:geomean,1);
      \coordinate (O1) at (rel axis cs:0,0);
      \coordinate (O2) at (rel axis cs:1,0);

      \draw [black,sharp plot,dashed] (A -| O1) -- (A -| O2);

      \legend{
        Copy-Patch x86-64,
        TPDE x86-64,
        TPDE AArch64
      }

    \end{axis}
  \end{tikzpicture}
}
  \caption{Run-time speedup}
  \label{fig:llvm-rt}
\end{subfigure}
  \caption{Compile- and Run-time speedup normalized to LLVM \texttt{-O0} on SPECint 2017 with unoptimized LLVM-IR.
  Compile-time is back-end time, excluding front-end and required LLVM-IR passes.}
\end{figure}
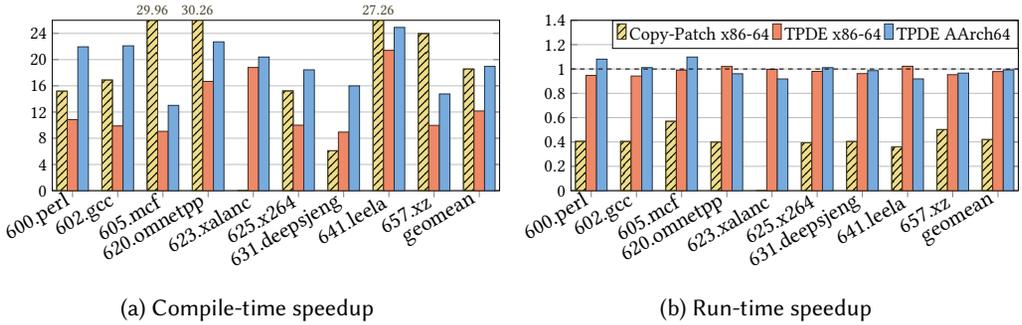

\begin{figure}
\begin{minipage}{.49\linewidth}
\resizebox{\linewidth}{!}{
  \begin{tikzpicture}
    \begin{axis}[
      xbar stacked,
      xmin=0,
      xmax=100,
      enlarge y limits=0.5,
      symbolic y coords={TPDE,Complete},
      ytick=data,
      legend cell align=left,
      legend style={
        area legend,
        at={(0.42,1)},
        anchor=south,
        legend columns = 3,
        draw=none,
      },
      bar width=15pt,
      cycle list={%
        TLorange!30!black,fill=TLorange,mark=none\\%
        TLlightblue!30!black,fill=TLlightblue,mark=none\\%
        TLpink!30!black,fill=TLpink,mark=none\\%
        TLlightyellow!30!black,fill=TLlightyellow,mark=none\\%
        TLlightcyan!30!black,fill=TLlightcyan,mark=none\\%
        TLpear!30!black,fill=TLpear,mark=none\\%
        TLmint!30!black,fill=TLmint,mark=none\\%
        TLolive!30!black,fill=TLolive,mark=none\\%
        TLpalegray!30!black,fill=TLpalegray,mark=none\\%
        TLlightblue!30!black,fill=TLlightblue,mark=none\\%
      },
      height=3.7cm,
      width=9cm,
      xlabel=Percent
      ]

      \draw [thick] (axis cs:98.5,{[normalized]0.75}) -- (axis cs:0,{[normalized]0.25});
      \draw [thick] (axis cs:100,{[normalized]0.75}) -- (axis cs:100,{[normalized]0.25});

      \addplot coordinates {(98.8,Complete)(0,TPDE)}; %
      \addplot coordinates {(1.2,Complete)(0,TPDE)}; %

      \addplot coordinates {(0,Complete)(13.58,TPDE)};%
      \addplot coordinates {(0,Complete)(12.71,TPDE)};%
      \addplot coordinates {(0,Complete)(48.81,TPDE)};%
      \addplot coordinates {(0,Complete)(24.9,TPDE)};%

      \coordinate (c1) at (axis cs:98.8,Complete);
      \coordinate (c2) at (axis cs:0,TPDE);

      \legend{
        Frontend (Clang),Backend (TPDE),
Preparation Pass,
Analysis Pass,
CodeGen Pass,
Miscellaneous
}

    \end{axis}
  \end{tikzpicture}
  }
  \caption{Time distribution when compiling all SPECint 2017 benchmarks (\texttt{-O0}). Miscellaneous is mostly measurement overhead and object file writing.}
  \label{fig:llvm:ct-breakup}
\end{minipage}
\hfill
\begin{minipage}{.49\linewidth}
\resizebox{\linewidth}{!}{
  \begin{tikzpicture}
    \begin{axis}[
      ybar=0pt,
      ymin=0,
      ymax=3.3,
      bar width=7pt,
      ymajorgrids=true,
      ytick distance=0.5,
      symbolic x coords = {
        600.perl,
        602.gcc,
        605.mcf,
        620.omnetpp,
        623.xalanc,
        625.x264,
        631.deepsjeng,
        641.leela,
        657.xz,
        geomean
      },
      enlarge x limits=0.05,
      xtick=data,
      xtick pos=bottom,
      cycle list={%
        TLlightyellow!30!black,fill=TLlightyellow,mark=none,postaction={pattern=north east lines}\\%
        TLorange!30!black,fill=TLorange,mark=none\\%
        TLlightblue!30!black,fill=TLlightblue,mark=none\\%
        TLpink!30!black,fill=TLpink,mark=none\\%
        TLlightyellow!30!black,fill=TLlightyellow,mark=none\\%
        TLlightcyan!30!black,fill=TLlightcyan,mark=none\\%
        TLolive!30!black,fill=TLolive,mark=none\\%
        TLmint!30!black,fill=TLmint,mark=none\\%
        TLpear!30!black,fill=TLpear,mark=none\\%
        TLpalegray!30!black,fill=TLpalegray,mark=none\\%
        TLlightblue!30!black,fill=TLlightblue,mark=none\\%
      },
      xticklabel style={rotate=30,anchor=east,align=center,font=\LARGE},
      yticklabel style={font=\Large},
      legend style={font=\Large,legend columns=-1,
        at={(1,1)},
        anchor=north east,},
      legend image code/.code={
        \draw [#1] (0cm,-0.1cm) rectangle (0.2cm,0.25cm); },
      nodes near coords boundcheck,
      width=12cm,
      height=5.2cm,
      ]

      \addplot coordinates {
        (600.perl,4.117595920742534)
        (602.gcc,4.388036094296581)
        (605.mcf,4.3019041156041675)
        (620.omnetpp,4.370927000481191)
        (623.xalanc,0)
        (625.x264,4.912533823492402)
        (631.deepsjeng,4.425515907136716)
        (641.leela,4.351628398924684)
        (657.xz,4.713927781871776)
        (geomean,4.441774922565391)
      };
      \addplot coordinates {
        (600.perl,1.28)
        (602.gcc,1.33)
        (605.mcf,1.28)
        (620.omnetpp,1.62)
        (623.xalanc,1.77)
        (625.x264,1.27)
        (631.deepsjeng,1.26)
        (641.leela,1.94)
        (657.xz,1.30)
        (geomean,1.43)
      };
      \addplot coordinates {
        (600.perl,1.09)
        (602.gcc,1.23)
        (605.mcf,1.15)
        (620.omnetpp,2.09)
        (623.xalanc,2.39)
        (625.x264,1.24)
        (631.deepsjeng,1.13)
        (641.leela,2.64)
        (657.xz,1.30)
        (geomean,1.49)
      };

      \coordinate (A) at (axis cs:geomean,1);
      \coordinate (O1) at (rel axis cs:0,0);
      \coordinate (O2) at (rel axis cs:1,0);

      \draw [black,sharp plot,dashed] (A -| O1) -- (A -| O2);

      \legend{
        Copy-Patch x86-64,
        TPDE x86-64,
        TPDE AArch64
      }

    \end{axis}
  \end{tikzpicture}
}
  \caption{Size of \texttt{.text} section of TPDE-generated code relative to LLVM-generated code on \texttt{-O0} IR.}
  \label{fig:llvm:codesize}
\end{minipage}
\end{figure}

\subsubsection{Results (Unoptimized IR)}
\autoref{fig:llvm-ct} shows the compile-time speedup over the LLVM \texttt{-O0} back-end when compiling unoptimized LLVM-IR. TPDE can generate code 8--24x faster compared to LLVM.
On AArch64, the compile-time improvements are larger (geomean: 18.96x) than on x86-64 (geomean: 12.15x), which is primarily caused by LLVM using the GlobalISel instruction selector by default, which is significantly slower than FastISel~\cite{engelke2024compile}.
TPDE is still substantially slower compared to the copy-and-patch compiler on x86-64 (which is geomean 18.6x faster than LLVM \texttt{-O0}) due to substantially more bookkeeping and explicit encoding of individual instructions. The copy-and-patch compiler is primarily limited by mapping LLVM-IR instructions to their templates, materializing constants, and generating moves to the registers expected by the templates.
In terms of run-time performance (cf. \autoref{fig:llvm-rt}), our generated code has a similar performance to LLVM ($\pm9\%$). %
The copy-and-patch-generated code is substantially slower (geomean: 2.38x slowdown) due to the huge amount of moves and spills caused by fixed registers used by the templates and the lack of a liveness analysis.

\autoref{fig:llvm:ct-breakup} shows the time distribution between Clang's front-end and TPDE as well the distribution of the passes within TPDE.
Within the entire compilation including the front-end, with TPDE only 2\% of the time are spent inside the back-end,
compared to 15\% with the default LLVM back-end (average speedup of end-to-end compilation: 17\%). This ratio is especially low for C++ programs,
where the front-end takes the largest portion of time. %
Of the time spent in TPDE, the largest part is code generation (49\%) followed by the LLVM preparation pass (14\%), which is caused by LLVM's data structured being comparably expensive to traverse and modify. The analysis pass (12\%) takes a comparably small amount of time.  %

We also measured the resulting code size for all binaries, \autoref{fig:llvm:codesize} shows the results, indicating a geomean increase of 43\% (x86-64) and 49\% (AArch64).
This increase is largely attributable to the substantially larger prologues/epilogues, which always reserve space for saving/restoring all callee-saved registers.
While this overhead is around 20-50\% for C-based benchmarks,
the code size increase is significantly higher for C++-based benchmarks. We currently generate all code into
a single text section and generate weak symbols instead of comdat sections for inline functions, %
preventing the linker from removing duplicate definitions. %
However, there is no structural problem implementing this behavior
into TPDE, however, the extra bookkeeping is likely to have a small impact on compile-time.
In contrast, the code from the copy-and-patch compiler is significantly larger than the LLVM \texttt{-O0} code (geomean: 4.44x), again primarily caused by the large amount of value moves.

\begin{figure}
\centering
\begin{subfigure}[b]{0.49\textwidth}
\resizebox{\textwidth}{!}{
  \begin{tikzpicture}
    \begin{axis}[
      ymode=log,
      ybar=0pt,
      ymin=0,
      ymax=120,
      ymajorgrids=true,
      yminorgrids=true,
      bar width=6pt,
      symbolic x coords = {
        600.perl,
        602.gcc,
        605.mcf,
        620.omnetpp,
        623.xalanc,
        625.x264,
        631.deepsjeng,
        641.leela,
        657.xz,
        geomean
      },
      enlarge x limits=0.05,
      xtick=data,
      xtick pos=bottom,
      cycle list={%
        TLpear!30!black,fill=TLpear,mark=none,postaction={pattern=crosshatch dots}\\%
        TLorange!30!black,fill=TLorange,mark=none\\%
        TLlightcyan!30!black,fill=TLlightcyan,mark=none,postaction={pattern=crosshatch dots}\\%
        TLlightblue!30!black,fill=TLlightblue,mark=none\\%
        TLorange!30!black,fill=TLorange,mark=none\\%
        TLlightblue!30!black,fill=TLlightblue,mark=none\\%
        TLpink!30!black,fill=TLpink,mark=none\\%
        TLlightyellow!30!black,fill=TLlightyellow,mark=none\\%
        TLlightcyan!30!black,fill=TLlightcyan,mark=none\\%
        TLolive!30!black,fill=TLolive,mark=none\\%
        TLmint!30!black,fill=TLmint,mark=none\\%
        TLpear!30!black,fill=TLpear,mark=none\\%
        TLpalegray!30!black,fill=TLpalegray,mark=none\\%
        TLlightblue!30!black,fill=TLlightblue,mark=none\\%
      },
      xticklabel style={rotate=30,anchor=east,align=center,font=\LARGE},
      yticklabel style={font=\Large},
      legend style={font=\large,legend columns=-1,
        at={(0,1)},
        anchor=north west,},
      width=12cm,
      height=5.5cm,
      ]

      \addplot table[x=benchmark,y expr=\thisrow{o1-x86_64-clang}/\thisrow{o1ir-x86_64-clang},col sep=comma] {bench-res/parsed-spec-ct};
      \addplot table[x=benchmark,y expr=\thisrow{o1-x86_64-clang}/\thisrow{o1ir-x86_64-tpde},col sep=comma] {bench-res/parsed-spec-ct};
      \addplot table[x=benchmark,y expr=\thisrow{o1-aarch64-clang}/\thisrow{o1ir-aarch64-clang},col sep=comma] {bench-res/parsed-spec-ct};
      \addplot table[x=benchmark,y expr=\thisrow{o1-aarch64-clang}/\thisrow{o1ir-aarch64-tpde},col sep=comma] {bench-res/parsed-spec-ct};

      \coordinate (A) at (axis cs:geomean,1);
      \coordinate (O1) at (rel axis cs:0,0);
      \coordinate (O2) at (rel axis cs:1,0);

    \end{axis}
  \end{tikzpicture}
}
  \caption{Compile-time speedup (log scale)}
  \label{fig:llvm-ct-opt}
\end{subfigure}
\begin{subfigure}[b]{0.49\textwidth}
\resizebox{\textwidth}{!}{
  \begin{tikzpicture}
    \begin{axis}[
      ybar=0pt,
      ymax=1.4,
      ymin=0,
      ymajorgrids=true,
      ytick distance=.2,
      bar width=6pt,
      symbolic x coords = {
        600.perl,
        602.gcc,
        605.mcf,
        620.omnetpp,
        623.xalanc,
        625.x264,
        631.deepsjeng,
        641.leela,
        657.xz,
        geomean
      },
      enlarge x limits=0.05,
      xtick=data,
      xtick pos=bottom,
      cycle list={%
        TLpear!30!black,fill=TLpear,mark=none,postaction={pattern=crosshatch dots}\\%
        TLorange!30!black,fill=TLorange,mark=none\\%
        TLlightcyan!30!black,fill=TLlightcyan,mark=none,postaction={pattern=crosshatch dots}\\%
        TLlightblue!30!black,fill=TLlightblue,mark=none\\%
        TLpink!30!black,fill=TLpink,mark=none\\%
        TLlightyellow!30!black,fill=TLlightyellow,mark=none\\%
        TLlightcyan!30!black,fill=TLlightcyan,mark=none\\%
        TLolive!30!black,fill=TLolive,mark=none\\%
        TLmint!30!black,fill=TLmint,mark=none\\%
        TLpear!30!black,fill=TLpear,mark=none\\%
        TLpalegray!30!black,fill=TLpalegray,mark=none\\%
        TLlightblue!30!black,fill=TLlightblue,mark=none\\%
      },
      xticklabel style={rotate=30,anchor=east,align=center,font=\LARGE},
      yticklabel style={font=\Large},
      legend style={font=\Large,legend columns=2,
        at={(1,1)},
        anchor=north east,},
      legend image code/.code={
        \draw [#1] (0cm,-0.1cm) rectangle (0.2cm,0.25cm); },
      width=12cm,
      height=5.5cm,
      ]

      \addplot table[x=benchmark,y expr=\thisrow{o1-x86_64-clang}/\thisrow{o1ir-x86_64-clang},col sep=comma] {bench-res/parsed-spec-rt};
      \addplot table[x=benchmark,y expr=\thisrow{o1-x86_64-clang}/\thisrow{o1ir-x86_64-tpde},col sep=comma] {bench-res/parsed-spec-rt};
      \addplot table[x=benchmark,y expr=\thisrow{o1-aarch64-clang}/\thisrow{o1ir-aarch64-clang},col sep=comma] {bench-res/parsed-spec-rt};
      \addplot table[x=benchmark,y expr=\thisrow{o1-aarch64-clang}/\thisrow{o1ir-aarch64-tpde},col sep=comma] {bench-res/parsed-spec-rt};

      \coordinate (A) at (axis cs:geomean,1);
      \coordinate (O1) at (rel axis cs:0,0);
      \coordinate (O2) at (rel axis cs:1,0);

      \draw [black,sharp plot,dashed] (A -| O1) -- (A -| O2);

      \legend{
        LLVM -O0 x86-64,
        TPDE x86-64,
        LLVM -O0 AArch64,
        TPDE AArch64
      }

    \end{axis}
  \end{tikzpicture}
}
  \caption{Run-time speedup}
  \label{fig:llvm-rt-opt}
\end{subfigure}
  \caption{Compile- and Run-time speedup normalized to the LLVM \texttt{-O1} back-end on SPECint 2017 with optimized (\texttt{-O1}) LLVM-IR.
  Compile-time is back-end time, excluding front-end and required LLVM-IR passes.}
\end{figure}
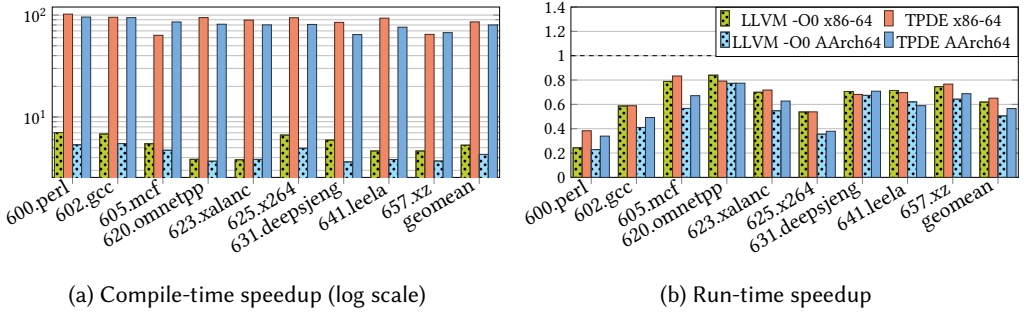

\subsubsection{Results (Optimized IR)}
\autoref{fig:llvm-ct-opt} shows the compile-time speedup over the LLVM \texttt{-O1} back-end when compiling optimized (\texttt{-O1}) LLVM-IR. TPDE achieves a geomean speedup in compile-time of 85.8x/80.0x over LLVM \texttt{-O1} and 16.2x/18.7x over LLVM \texttt{-O0}. Therefore, also for optimized input IR, TPDE achieves similar compilation performance as for unoptimized IR.
In terms of run-time performance (cf. \autoref{fig:llvm-rt-opt}), our generated code often has a slightly better performance compared to the code generated by the LLVM \texttt{-O0} back-end (geomean x86-64: 5\%, AArch64: 11\%). Compared to the optimizing back-end, however, TPDE-generated code is substantially slower (geomean x86-64: 1.54x, AArch64: 1.77x), primarily because LLVM in this configuration uses a much better register allocator and instruction selector.  %

\subsection{Discussion}
The results show that TPDE allows implementing a fast back-end for a general-purpose IR like LLVM-IR with reasonable effort. The compilation times are substantially faster than LLVM's due to the single-pass code generation approach in contrast to LLVM's multitude of IR conversions and rewrites on data structures that are expensive to modify (e.g., Machine IR). Additionally, selecting algorithms and data structures with a focus on performance shows its advantage. Although TPDE is not as fast as a copy-and-patch-based compiler, the compile-times are in the same order of magnitude and the quality of the generated code is on-par with LLVM \texttt{-O0} in terms of performance.
As expected, TPDE-generated code is substantially slower than code produced by LLVM \texttt{-O1}. Nonetheless, further improvements regarding instruction selection might be possible by utilizing more information from the Machine IR snippets, this is left as future work.

\section{Case Study: Compiling WebAssembly}
\label{sec:cranelift}
Wasmtime~\cite{wasmtime} is a WebAssembly run-time which performs JIT compilation using either the single-pass back-end Winch
or the multi-pass back-end Cranelift. The latter still aims to be fast but can optionally perform optimizations on its SSA-IR called CLIF,
use a single-pass or backtracking register allocator, and do more elaborate instruction selection.
Since TPDE requires an SSA-IR to compile, we created a TPDE back-end for CLIF.
In contrast to LLVM-IR, CLIF uses block arguments instead of $\phi$-nodes, only supports a limited set of scalar and vector types,
and models stack slots separately, referencing them with explicit instructions. Additionally, CLIF also supports values which transparently alias
other values. A more detailed description of CLIF can be found in the Cranelift IR reference~\cite{clifdoc}.

\subsection{Implementation}
\subsubsection{IR Adapter}
As Cranelift is written in Rust, the adaptor needs to provide a cross-language interface to the IR.
Most information can be gathered easily from CLIF.
Since CLIF treats global values, stack slots, and arguments differently from normal IR values, the adaptor has to
create dummy values for them as the framework references them as regular values.

However, as CLIF uses block arguments and TPDE currently does not support multi-edges with different values, the adaptor has to place empty basic blocks on such edges in a preparation pass.
For more efficient iteration over block successors through C++ code, the adaptor also creates arrays of block successors.
In the same pass, also value aliases are eliminated, simplifying these cases during code generation.

\subsubsection{Compilation}
Many instructions can be implemented using snippet encoders with similar exceptions as LLVM.
Additionally, since many instructions share semantics with LLVM instructions with a smaller set of supported types, most snippets and instruction compilers can be reused from
the LLVM implementation in simplified form.

In contrast to LLVM, CLIF supports constants as results from special instructions and has separate instructions to generate pointers
to stack slots,
which need special handling so that they can be fused into other instructions.

\subsubsection{Code Complexity}
In total, the Cranelift back-end using TPDE consists of roughly 4k lines of code excluding blanks and comments which are specific to the Cranelift implementation,
out of which 700 are architecture-specific to x86-64 and roughly 1.6k are glue code to exchange data between C++ and Rust.
However, the back-end currently does not support any vector operations which significantly lowers the necessary complexity.

\subsection{Evaluation}
\subsubsection{Setup}
We evaluate the performance of the Cranelift back-end by measuring compile- and run-time on the three default benchmarks
in Wasmtime's own benchmark suite Sightglass~\cite{sightglass} and PolyBench~\cite{PolyBenchC}.
We compare this against Cranelift with its backtracking and single pass register allocator, both without any IR optimizations, and Winch.
The benchmark machines are the same as in \autoref{sec:llvm:eval}, we report the speedup using an average of 10
compilations and 5 executions.

\subsubsection{Results}
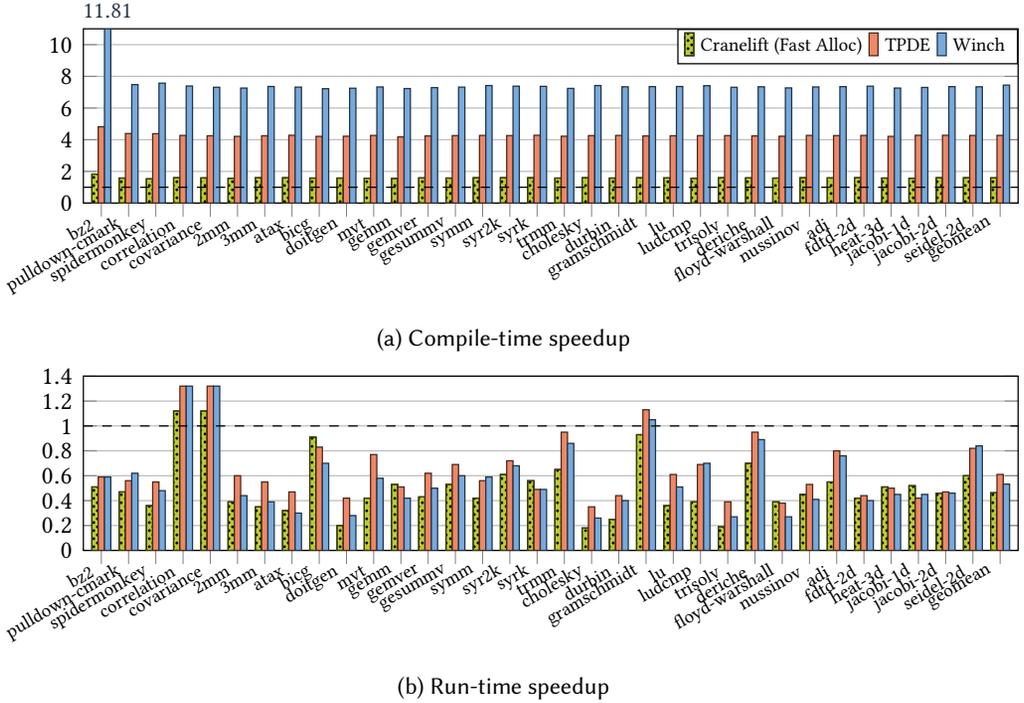
\begin{figure}
\centering
\begin{subfigure}[b]{\textwidth}
\resizebox{\textwidth}{!}{
  \begin{tikzpicture}
    \begin{axis}[
      ybar=0pt,
      ymin=0,
      ymax=11,
      ymajorgrids=true,
      ytick distance=2,
      bar width=2pt,
      symbolic x coords = {
        bz2,
        pulldown-cmark,
        spidermonkey,
        correlation,
        covariance,
        2mm,
        3mm,
        atax,
        bicg,
        doitgen,
        mvt,
        gemm,
        gemver,
        gesummv,
        symm,
        syr2k,
        syrk,
        trmm,
        cholesky,
        durbin,
        gramschmidt,
        lu,
        ludcmp,
        trisolv,
        deriche,
        floyd-warshall,
        nussinov,
        adi,
        fdtd-2d,
        heat-3d,
        jacobi-1d,
        jacobi-2d,
        seidel-2d,
        geomean
      },
      enlarge x limits=0.02,
      xtick=data,
      xtick pos=bottom,
      cycle list={%
        TLpear!30!black,fill=TLpear,mark=none,postaction={pattern=crosshatch dots}\\%
        TLorange!30!black,fill=TLorange,mark=none\\%
        TLlightblue!30!black,fill=TLlightblue,mark=none\\%
        TLpink!30!black,fill=TLpink,mark=none\\%
        TLlightyellow!30!black,fill=TLlightyellow,mark=none\\%
        TLlightcyan!30!black,fill=TLlightcyan,mark=none\\%
        TLolive!30!black,fill=TLolive,mark=none\\%
        TLmint!30!black,fill=TLmint,mark=none\\%
        TLpear!30!black,fill=TLpear,mark=none\\%
        TLpalegray!30!black,fill=TLpalegray,mark=none\\%
        TLlightblue!30!black,fill=TLlightblue,mark=none\\%
      },
      xticklabel style={rotate=30,anchor=east,align=center,font=\tiny},
      yticklabel style={font=\footnotesize},
      legend style={thin,font=\tiny,legend columns=-1,inner ysep=0.5pt, inner xsep=2pt,
        at={(1,1)},
        anchor=north east,},
        legend image code/.code={
        \draw [#1] (0cm,-0.1cm) rectangle (0.1cm,0.15cm); },
      width=12cm,
      height=3.5cm,
      nodes near coords boundcheck,
      every node near coord/.append style={font=\scriptsize},
      ]

      \addplot table[x=bench,y=ct,col sep=comma] {bench-res/res-wasmtime-cfa};
      \addplot table[x=bench,y=ct,col sep=comma] {bench-res/res-wasmtime-tpde};
      \addplot table[x=bench,y=ct,col sep=comma] {bench-res/res-wasmtime-winch};

      \coordinate (A) at (axis cs:geomean,1);
      \coordinate (O1) at (rel axis cs:0,0);
      \coordinate (O2) at (rel axis cs:1,0);

      \draw [black,sharp plot,dashed] (A -| O1) -- (A -| O2);

      \legend{
        Cranelift (Fast Alloc),
        TPDE,
        Winch
      }

    \end{axis}
  \end{tikzpicture}
}

  \caption{Compile-time speedup}
  \label{fig:cranelift-ct}
\end{subfigure}
\begin{subfigure}[b]{\textwidth}
\resizebox{\textwidth}{!}{
  \begin{tikzpicture}
    \begin{axis}[
      ybar=0pt,
      ymax=1.4,
      ymin=0,
      ymajorgrids=true,
      ytick distance=0.2,
      bar width=2pt,
      symbolic x coords = {
        bz2,
        pulldown-cmark,
        spidermonkey,
        correlation,
        covariance,
        2mm,
        3mm,
        atax,
        bicg,
        doitgen,
        mvt,
        gemm,
        gemver,
        gesummv,
        symm,
        syr2k,
        syrk,
        trmm,
        cholesky,
        durbin,
        gramschmidt,
        lu,
        ludcmp,
        trisolv,
        deriche,
        floyd-warshall,
        nussinov,
        adi,
        fdtd-2d,
        heat-3d,
        jacobi-1d,
        jacobi-2d,
        seidel-2d,
        geomean
      },
      enlarge x limits=0.02,
      xtick=data,
      xtick pos=bottom,
      cycle list={%
        TLpear!30!black,fill=TLpear,mark=none,postaction={pattern=crosshatch dots}\\%
        TLorange!30!black,fill=TLorange,mark=none\\%
        TLlightblue!30!black,fill=TLlightblue,mark=none\\%
        TLpink!30!black,fill=TLpink,mark=none\\%
        TLlightyellow!30!black,fill=TLlightyellow,mark=none\\%
        TLlightcyan!30!black,fill=TLlightcyan,mark=none\\%
        TLolive!30!black,fill=TLolive,mark=none\\%
        TLmint!30!black,fill=TLmint,mark=none\\%
        TLpear!30!black,fill=TLpear,mark=none\\%
        TLpalegray!30!black,fill=TLpalegray,mark=none\\%
        TLlightblue!30!black,fill=TLlightblue,mark=none\\%
      },
      xticklabel style={rotate=30,anchor=east,align=center,font=\tiny},
      yticklabel style={font=\footnotesize},
      legend style={thin,font=\tiny,legend columns=-1,inner ysep=0.5pt, inner xsep=2pt,
        at={(1,1)},
        anchor=north east,},
        legend image code/.code={
        \draw [#1] (0cm,-0.1cm) rectangle (0.1cm,0.15cm); },
      width=12cm,
      height=3.5cm,
      ]

      \addplot table[x=bench,y=rt,col sep=comma] {bench-res/res-wasmtime-cfa};
      \addplot table[x=bench,y=rt,col sep=comma] {bench-res/res-wasmtime-tpde};
      \addplot table[x=bench,y=rt,col sep=comma] {bench-res/res-wasmtime-winch};

      \coordinate (A) at (axis cs:geomean,1);
      \coordinate (O1) at (rel axis cs:0,0);
      \coordinate (O2) at (rel axis cs:1,0);

      \draw [black,sharp plot,dashed] (A -| O1) -- (A -| O2);

    \end{axis}
  \end{tikzpicture}
}

  \caption{Run-time speedup}
  \label{fig:cranelift-rt}
\end{subfigure}
  \caption{Compile- and Run-time speedup normalized to Cranelift with the default register allocator on three Sightglass and all PolyBench benchmarks.
  Compile-time includes translation into CLIF and linking.}
  \label{fig:cranelift:perf}
\end{figure}

\autoref{fig:cranelift:perf} shows the results.
The TPDE-based back-end compiles 4.27x faster than Cranelift and 2.68x faster than Cranelift with its fast register allocator, but is 1.74x slower than Winch. There are two primary reasons for this: first, 37\% of the time are spent on translating WebAssembly to CLIF; and second, this translation already constructs SSA form for all variables, which is not needed by TPDE and also produces many trivially removable $\phi$-nodes.
Constructing a more light-weight IR directly from WebAssemebly could significantly close the gap to Winch's compile times, as only 40\% of the time are spent inside TPDE on IR analysis and code generation.

The run-time performance of
TPDE-generated code is faster than both Winch and Cranelift with its fast register
allocator (1.14x and 1.31x respectively), but 1.64x slower than Cranelift with its default backtracking register allocator.
This shows that a more sophisticated register allocation heuristic is likely to substantially improve the run-time performance.

\section{Case Study: Compiling Umbra IR}
\label{sec:umbra}
Umbra~\cite{neumann2020umbra} is a compiling database system using JIT compilation to execute SQL queries efficiently.
As queries are not always known ahead-of-time, the query latency consisting of compilation and execution time should be as low as possible.
To easily allow switching between different compilation back-ends, Umbra first generates all code into its custom IR~\cite{neumann2020umbra,kersten2021tidy}, which makes heavy use of dense data structures for better cache utilization and allow for fast iteration when lowering further for compilation.
The IR is in SSA form, inspired by LLVM-IR, %
but only supports a small set of data types composed of 8/16/32/64/128-bit integers, pointers, double precision
floating-point and \texttt{data128}, which consists of two 64-bit integers and is often used for strings.
Additionally, Umbra IR has several instructions to briefly express frequently occurring operations, %
for example, the \texttt{ssubtrap} instruction, which performs a signed subtraction, calls a trap function if an overflow occurs, and otherwise returns the result of the subtraction. A more detailed description of Umbra IR can be found in~\cite{kersten2021tidy}.

To achieve optimal execution time for queries of different workload sizes, Umbra has multiple back-ends with different performance characteristics.
For optimized code generation, Umbra IR can be translated to LLVM-IR, which is
then transformed using LLVM's optimization passes and compiled using the optimizing LLVM back-end. Due to long compile times, this back-end is only used
for long-running queries where overhead of optimization does not negate the benefit of more efficient code.
For fast code generation, Umbra supports using the non-optimizing LLVM \texttt{-O0} pipeline.
To further reduce latency, the default back-end is a custom-written \emph{DirectEmit} back-end which compiles Umbra IR in two-passes to machine code while
achieving a run-time performance typically better than the LLVM baseline.
However, this back-end has a very high code complexity and is extremely platform-dependent; for its port to AArch64,
the code had to be largely duplicated with little code reuse.
In addition to that, the expansion of all instructions to machine code has to be handwritten
which makes this back-end hard to maintain and port to new architectures.

Therefore, we wrote an Umbra back-end using TPDE with the goal of compiling code as fast as DirectEmit, producing a similar code quality, while at the same time reducing the amount of platform-dependent code and reducing code complexity in general.

\subsection{Implementation}

\subsubsection{IR Adapter}
Almost all information required by the adapter interface can be directly gathered from Umbra's IR data structures
with little to no overhead.
Since Umbra IR already has unique per-function IDs for instructions and blocks, we expose these as value and block references to the framework.
Unlike for our LLVM back-end, we do not need a preparation pass, as all instructions can be translated in a straight-forward manner. Numbering of globals can be performed lazily during code generation.

\subsubsection{Compilation}

Most instructions can be implemented using snippet encoders with the same exceptions as for LLVM. Additionally, since
several instructions work similarily to their LLVM counterpart or are simple combination of instructions, we could reuse many snippets from our LLVM compiler,
but, as possible data types are more restricted in Umbra IR,
their implementation could often be simplified.
As Umbra only uses JIT compilation, the addresses of called functions and referenced globals are known during compilation.
Therefore,
we can simply hardcode addresses of symbols into the generated code.%

\subsubsection{Code Complexity}
In total, the Umbra IR back-end using TPDE consists of 3.6k lines of code excluding blanks and comments, out of which 1.6k are target-specific for both, x86-64 and AArch64, combined. This is significantly less
than the implementation of DirectEmit with 11k lines of code for both AArch64 and x86-64. This makes our implementation substantially closer in length
to the LLVM back-end, which consists of 2.3k lines, mostly for the translation of IR semantics.

\subsection{Evaluation}
\subsubsection{Setup}
We evaluate the performance of our back-end by measuring the compile- and run-time on the TPC-DS~\cite{tpcds} benchmark using scale factor 1
and compare it against DirectEmit and LLVM with the regular LLVM back-ends as well as our TPDE-based LLVM back-end (TPDE-LLVM); the benchmark machines are the same as in \autoref{sec:llvm:eval}, we report the average of 20 compilation and execution runs.

\subsubsection{Results}
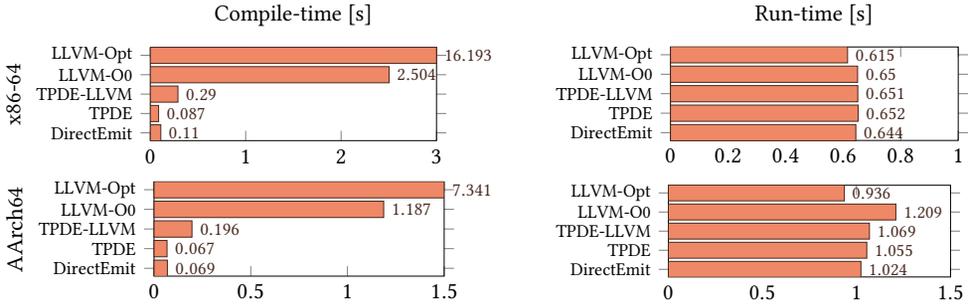
\begin{figure}
  \begin{subfigure}{0.49\textwidth}
\resizebox{\textwidth}{!}{
  \begin{tikzpicture}
    \begin{axis}[
      title={Compile-time [s]},
      ylabel={x86-64},
      xbar,
      xmin=0,
      xmax=3,
      symbolic y coords = {
        DirectEmit,
        TPDE,
        TPDE-LLVM,
        LLVM-O0,
        LLVM-Opt
      },
      ytick=data,
      cycle list={%
        TLorange!30!black,fill=TLorange,mark=none\\%
        TLlightblue!30!black,fill=TLlightblue,mark=none\\%
        TLpink!30!black,fill=TLpink,mark=none\\%
        TLlightyellow!30!black,fill=TLlightyellow,mark=none\\%
        TLlightcyan!30!black,fill=TLlightcyan,mark=none\\%
        TLolive!30!black,fill=TLolive,mark=none\\%
        TLmint!30!black,fill=TLmint,mark=none\\%
        TLpear!30!black,fill=TLpear,mark=none\\%
        TLpalegray!30!black,fill=TLpalegray,mark=none\\%
        TLlightblue!30!black,fill=TLlightblue,mark=none\\%
      },
      yticklabel style={align=center,font=\footnotesize},
      width=6cm,
      height=3cm,
      nodes near coords boundcheckx,
      every node near coord/.style={
        /pgf/number format/precision=3,
        /pgf/number format/fixed,
        font=\footnotesize,
      },
      bar width=7pt,
      ]

      \addplot+ table[x=ct,y=backend,col sep=comma] {bench-res/parsed-umbra-x86_64};
    \end{axis}
  \end{tikzpicture}
}
  \end{subfigure}
  \begin{subfigure}{0.445\textwidth}
\resizebox{\textwidth}{!}{
  \begin{tikzpicture}
    \begin{axis}[
      title={Run-time [s]},
      ylabel={\phantom{A}},
      xbar,
      xmin=0,
      xmax=1,
      symbolic y coords = {
        DirectEmit,
        TPDE,
        TPDE-LLVM,
        LLVM-O0,
        LLVM-Opt
      },
      ytick=data,
      cycle list={%
        TLorange!30!black,fill=TLorange,mark=none\\%
        TLlightblue!30!black,fill=TLlightblue,mark=none\\%
        TLpink!30!black,fill=TLpink,mark=none\\%
        TLlightyellow!30!black,fill=TLlightyellow,mark=none\\%
        TLlightcyan!30!black,fill=TLlightcyan,mark=none\\%
        TLolive!30!black,fill=TLolive,mark=none\\%
        TLmint!30!black,fill=TLmint,mark=none\\%
        TLpear!30!black,fill=TLpear,mark=none\\%
        TLpalegray!30!black,fill=TLpalegray,mark=none\\%
        TLlightblue!30!black,fill=TLlightblue,mark=none\\%
      },
      yticklabel style={align=center,font=\footnotesize},
      width=6cm,
      height=3cm,
      nodes near coords boundcheckx,
      every node near coord/.style={
        /pgf/number format/precision=3,
        /pgf/number format/fixed,
        font=\footnotesize,
      },
      bar width=7pt,
      ]

      \addplot+ table[x=rt,y=backend,col sep=comma] {bench-res/parsed-umbra-x86_64};
    \end{axis}
  \end{tikzpicture}
}
  \end{subfigure}

  \begin{subfigure}{0.49\textwidth}
\resizebox{\textwidth}{!}{
  \begin{tikzpicture}
    \begin{axis}[
      ylabel={AArch64},
      xbar,
      xmin=0,
      xmax=1.5,
      symbolic y coords = {
        DirectEmit,
        TPDE,
        TPDE-LLVM,
        LLVM-O0,
        LLVM-Opt
      },
      ytick=data,
      cycle list={%
        TLorange!30!black,fill=TLorange,mark=none\\%
        TLlightblue!30!black,fill=TLlightblue,mark=none\\%
        TLpink!30!black,fill=TLpink,mark=none\\%
        TLlightyellow!30!black,fill=TLlightyellow,mark=none\\%
        TLlightcyan!30!black,fill=TLlightcyan,mark=none\\%
        TLolive!30!black,fill=TLolive,mark=none\\%
        TLmint!30!black,fill=TLmint,mark=none\\%
        TLpear!30!black,fill=TLpear,mark=none\\%
        TLpalegray!30!black,fill=TLpalegray,mark=none\\%
        TLlightblue!30!black,fill=TLlightblue,mark=none\\%
      },
      yticklabel style={align=center,font=\footnotesize},
      width=6cm,
      height=3cm,
      nodes near coords boundcheckx,
      every node near coord/.style={
        /pgf/number format/precision=3,
        /pgf/number format/fixed,
        font=\footnotesize,
      },
      bar width=7pt,
      ]

      \addplot+ table[x=ct,y=backend,col sep=comma] {bench-res/parsed-umbra-aarch64};
    \end{axis}
  \end{tikzpicture}
}
  \end{subfigure}
  \begin{subfigure}{0.445\textwidth}
\resizebox{\textwidth}{!}{
  \begin{tikzpicture}
    \begin{axis}[
      ylabel={\phantom{AArch64}},
      xbar,
      xmin=0,
      xmax=1.5,
      symbolic y coords = {
        DirectEmit,
        TPDE,
        TPDE-LLVM,
        LLVM-O0,
        LLVM-Opt
      },
      ytick=data,
      cycle list={%
        TLorange!30!black,fill=TLorange,mark=none\\%
        TLlightblue!30!black,fill=TLlightblue,mark=none\\%
        TLpink!30!black,fill=TLpink,mark=none\\%
        TLlightyellow!30!black,fill=TLlightyellow,mark=none\\%
        TLlightcyan!30!black,fill=TLlightcyan,mark=none\\%
        TLolive!30!black,fill=TLolive,mark=none\\%
        TLmint!30!black,fill=TLmint,mark=none\\%
        TLpear!30!black,fill=TLpear,mark=none\\%
        TLpalegray!30!black,fill=TLpalegray,mark=none\\%
        TLlightblue!30!black,fill=TLlightblue,mark=none\\%
      },
      yticklabel style={align=center,font=\footnotesize},
      width=6cm,
      height=3cm,
      nodes near coords boundcheckx,
      every node near coord/.style={
        /pgf/number format/precision=3,
        /pgf/number format/fixed,
        font=\footnotesize,
      },
      bar width=7pt,
      ]

      \addplot+ table[x=rt,y=backend,col sep=comma] {bench-res/parsed-umbra-aarch64};
    \end{axis}
  \end{tikzpicture}
}
  \end{subfigure}

  \caption{Compile- and run-time accumulated over all TPC-DS queries at scale factor 1. Compilation is repeated 20 times for each query. Compile-time is code-generation time, excluding query plan and IR construction.}
\label{fig:umbra:perf}
\end{figure}

\autoref{fig:umbra:perf} shows the results. TPDE can generate code that is comparable to code produced by DirectEmit while having a similar compile-time performance.
The speedup in compilation-time over LLVM is especially large, because Umbra IR has to be translated into LLVM IR first.
The LLVM back-end written with TPDE achieves a substantial speedup already; however, the cost of translation to and preprocessing of LLVM-IR becomes visible.

The run-time performance increase compared to LLVM on AArch64 is caused by generating more optimized instruction sequences for complex IR instructions, while for the LLVM back-end, these have to be separated into simpler LLVM-IR instructions first, which are not fused by LLVM at \texttt{-O0}.

\subsection{Discussion}
The results show that our IR adapter and our generalized framework introduce only minimal additional overhead compared to the DirectEmit back-end, which is explicitly written for this specific IR. TPDE enables to write a back-end with similar compile-time and run-time performance with substantially less effort and a much smaller amount of architecture-specific code.
Although our TPDE-based LLVM back-end already provides substantially faster compilation over LLVM \texttt{-O0}, the extra IR translation has a measurable cost. This demonstrates that our approach of adapting the framework to existing IRs can significantly reduce the compilation latency.

\section{Related Work}
\label{sec:rltdwork}

\paragraph{IR-Independent Compilation Approaches}
To the best of our knowledge, a largely IR-independent compiler back-end framework that directly interfaces with existing IRs has not been proposed before.

A long-standing scheme for fast compilation consists of pre-compiling templates to machine code with a standard compiler and then simply concatenating them to quickly generate machine code, optionally with patching constants~\cite{ertl2004retargeting}. One early implementer of this approach was QEMU~\cite{bellard2005qemu}, which identified patch points through relocations, but later moved to a more sophisticated code generation approach. More recently, the approach regained traction under the name ``copy-and-patch''~\cite{xu2021copy}, which was later adapted for compiling MLIR~\cite{drescher2024fast} and Python~\cite{python2024pep744}.
A fundamental limitation of the approach is the lack of adaptability: as the machine code templates are combined in binary format, there is no way to change registers, replace register operands with immediates, or use different addressing modes in the machine code. Although some of these limitations can be partially alleviated by precompiling multiple variants, this is not practically possible for the general case as the number of required templates would be huge. Furthermore, the lack of a liveness analysis results in very frequent register moves and stack spills/reloads, causing a substantial slowdown in run-time performance. In contrast, our generated snippet encoders utilize information from LLVM's Machine IR to dynamically manage registers and flexibly morph the machine code to the actual operands, resulting in a performance that is on-par with LLVM \texttt{-O0}, while still having a very low code generation time due to our single-pass approach. Moreover, TPDE also has built-in support for $\phi$-nodes, dynamically-sized stack allocations, and C++ exceptions, which are not easily implementable in a template-based code generation approach.

AsmJit~\cite{asmjit} provides a low-level abstraction for generating x86 and AArch64 machine code and also a more high-level API, where a user encodes instructions referencing virtual registers, delegating register allocation to the framework. This is implemented by materializing all to-be-generated instructions in a custom in-memory IR, on which register allocation and code emission are performed as two separate passes. Our approach, in contrast, avoids the separate IR materialization by directly emitting machine code. %
Additionally, our framework is more high-level and targets compiling SSA code, especially by performing $\phi$-node elimination, and has built-in support for generating object files, unwind information, and code for C++ exception handling. Furthermore, our approach to generate snippet encoders from a high-level language provides an architecture-independent way to specify instruction semantics, allowing back-end writers to focus on operation semantics while still giving the flexibility to fine-tune the generated machine code where needed.

\paragraph{Fast Compilers for Fixed Code Representation}
Many runtime systems implement their own IR and, to reduce latency, implemented a custom back-end without reusing existing compiler frameworks and without an extra IR translation step. Examples include WebKit~\cite{webkit2016b3}, Wasmtime's Cranelift~\cite{byta2023cranelift}, and Umbra's DirectEmit back-end~\cite{kersten2021tidy}.
However, rolling custom IRs with low-latency compilation back-ends and porting these to all required architectures is a substantial effort. Furthermore, these IRs and compilers are often deeply embedded into large systems, preventing reuse in other projects. Even if a reuse were easily possible, this would nonetheless require writing an IR translator, which would unnecessarily increase compilation times.

With TPDE, we provide a highly efficient and adaptable compiler framework, which allows projects to keep their custom IRs as desired, but substantially reduce the effort of writing a code generator and ease portability to different platforms by specifying IR semantics in architecture-independent languages.

A key design decision shared by many fast compilers is to reduce the number of code transformations.
While optimizing frameworks like LLVM incrementally lower and rewrite the IR, fast compilers as implemented in V8~\cite{v8sparkplug,v8maglev} or WebKit JavaScriptCore~\cite{webkitjsc} run substantially fewer passes and their baseline compiler typically generates code in a single pass. While the lowest tier typically just concatenates simple code fragments or runtime calls to avoid the interpreter dispatch overhead, the mid-level tier of these systems (V8's Maglev~\cite{v8maglev}, JSC's DFG JIT~\cite{webkitjsc}) performs some lightweight analyses and optimizations while using fast algorithms like a greedy register allocator as separate passes.

Our approach in TPDE can be considered as a hybrid of the baseline and mid-level tier: we do single-pass code generation and do not perform changes on the IR, but still run a fast liveness and loop analysis to increase the quality of the generated code.

Another relevant concept is the preference of dense data structures like arrays and bit sets over hash maps/sets. For this purpose, IR values and blocks are often numbered. Performance-focused compilers also try to avoid unnecessary computations, for example, by not having an always-up-to-date use list for SSA values (e.g., JSC's B3, V8's Maglev, Go, Umbra IR) or by not using SSA form at all (e.g., JSC's DFG initially starts in non-SSA form, V8's Sparkplug). TPDE similarly makes heavy use of block and value numbers for fast lookups, but otherwise performs no IR changes at all.

Domain-specific compilers targeting, for example, WebAssembly~\cite{titzer2024whose} can achieve lower latency by designing the input format in a way that is easy to compile and making use of this structure in the baseline compiler. %

\section{Conclusion}
\label{sec:summary}

In this paper, we presented TPDE, a compiler back-end framework for fast machine code generation that adapts to existing IRs in SSA form.
To adapt the framework for their IR, a user has to provide two components: an IR adapter, which provides a canonical way of accessing IR data structures for the framework, and instruction compilers, which implement the actual semantics of the IR instructions.
For increased portability and ease-of-use, these instruction compilers can, optionally, largely be auto-generated from a higher-level language like C using LLVM's compilation pipeline. The machine instruction sequences are extracted from LLVM's Machine IR, which provides extensive meta-information and therefore permits more local optimizations.
For compilation, the framework performs just two passes: an analysis pass, which performs a liveness analysis of the defined IR values, and a code generation pass, which performs the combination of lowering to machine instructions, register allocation, and instruction encoding.

We used TPDE to build a three-pass compiler for the commonly used subset of LLVM-IR targeting x86-64 and AArch64, which can compile code 8--24x faster than the existing LLVM \texttt{-O0} back-end while achieving similar run-time performance.
Furthermore, we used TPDE to implement a compilation back-end for the Wasmtime WebAssembly runtime as well as the Umbra database system, adapting the framework to other existing IRs in the context of JIT compilation. For Umbra, our TPDE-based back-end was on-par with its highly-specialized direct code emission back-end while having a significantly lower implementation complexity.

\bibliographystyle{ACM-Reference-Format}
\bibliography{paper}

\end{document}